\documentclass[times, twocolumn]{aastex63}
\usepackage{graphicx, graphics}
\usepackage{CJK}
\usepackage{bbm}
\definecolor{cyan(process)}{rgb}{0.0, 0.72, 0.92}
\usepackage{amsmath, amssymb, bm}
\newcommand*{\QED}{\hfill\ensuremath{\blacksquare}}%
\usepackage[nodayofweek]{datetime}
\newdateformat{monthyearday}{%
  \THEYEAR\ \monthname[\THEMONTH] \twodigit{\THEDAY}}
\newcommand\numberthis{\addtocounter{equation}{1}\tag{\theequation}}
\newcommand{\uatnum}[1]{\href{http://vocabs.ands.org.au/repository/api/lda/aas/the-unified-astronomy-thesaurus/current/resource.html?uri=http://astrothesaurus.org/uat/#1}{#1}}

\received{2019 August 19}
\revised{2020 January 20} 
\accepted{2020 January 25}
\published{2020 March 31}

\shorttitle{NMF Data Imputation}
\shortauthors{Ren et al.}

\usepackage{romannum}

\begin{document}
\pagenumbering{arabic}
\begin{CJK*}{UTF8}{gbsn}
\title{Using Data Imputation for Signal Separation in High Contrast Imaging}
\author[0000-0003-1698-9696]{Bin Ren (任彬)}\email{ren@caltech.edu}
\affiliation{Department of Astronomy, California Institute of Technology, 1216 East California Boulevard, Pasadena, CA 91125, USA}
\affiliation{Department of Physics and Astronomy, The Johns Hopkins University, 3701 San Martin Drive, Baltimore, MD 21218, USA}
\affiliation{Department of Applied Mathematics and Statistics, The Johns Hopkins University, 3400 North Charles Street, Baltimore, MD 21218, USA}

\author{Laurent Pueyo}
\affiliation{Space Telescope Science Institute (STScI), 3700 San Martin Drive, Baltimore, MD 21218, USA}

\author[0000-0002-8382-0447]{Christine Chen} 
\affiliation{Space Telescope Science Institute (STScI), 3700 San Martin Drive, Baltimore, MD 21218, USA}
\affiliation{Department of Physics and Astronomy, The Johns Hopkins University, 3701 San Martin Drive, Baltimore, MD 21218, USA}

\author[0000-0002-9173-0740]{\'Elodie Choquet}
\affiliation{Aix Marseille Univ, CNRS, CNES, LAM, Marseille, France}

\author[0000-0002-1783-8817]{John H. Debes}
\affiliation{Space Telescope Science Institute (STScI), 3700 San Martin Drive, Baltimore, MD 21218, USA}

\author[0000-0002-5092-6464]{Gaspard Duch\^ene}
\affiliation{Astronomy Department, University of California, Berkeley, CA 94720, USA}
\affiliation{Universit\'e Grenoble-Alpes, CNRS Institut de Plan\'etologie et d'Astrophysique (IPAG), F-38000 Grenoble, France}

\author[0000-0002-1637-7393]{Fran\c{c}ois M\'enard} 
\affiliation{Universit\'e Grenoble-Alpes, CNRS Institut de Plan\'etologie et d'Astrophysique (IPAG), F-38000 Grenoble, France}

\author[0000-0002-3191-8151]{Marshall D. Perrin}
\affiliation{Space Telescope Science Institute (STScI), 3700 San Martin Drive, Baltimore, MD 21218, USA}

\begin{abstract}
To characterize circumstellar systems in high contrast imaging, the fundamental step is to construct a best point spread function (PSF) template for the non-circumstellar signals (i.e., star light and speckles) and separate it from the observation. With existing PSF construction methods, the circumstellar signals (e.g., planets, circumstellar disks) are unavoidably altered by over-fitting and/or self-subtraction, making forward modeling a necessity to recover these signals. We present a forward modeling--free solution to these problems with data imputation using sequential non-negative matrix factorization (DI-sNMF). DI-sNMF first converts this signal separation problem to a ``missing data'' problem in statistics by flagging the regions which host circumstellar signals as missing data, then attributes PSF signals to these regions. We mathematically prove it to have negligible alteration to circumstellar signals when the imputation region is relatively small, which thus enables precise measurement for these circumstellar objects. We apply it to simulated point source and circumstellar disk observations to demonstrate its proper recovery of them. We apply it to Gemini Planet Imager (GPI) $K1$-band observations of the debris disk surrounding HR~4796A, finding a tentative trend that the dust is more forward scattering as the wavelength increases. We expect DI-sNMF to be applicable to other general scenarios where the separation of signals is needed.
\end{abstract}


\section{Introduction}
High contrast imaging of circumstellar systems in visible and near infrared light offers direct spectroscopic and astrometric information on the planet or the distribution of the dust in the system. With the advancements in both instrumental design (e.g., Gemini/GPI: \citealp{macintosh08, macintosh14}; {VLT}/SPHERE: \citealp{beuzit08, beuzit19, vigan16}) and data post-processing (e.g., \citealp{lafreniere07, soummer12, amara12, milli12, pueyo12, pueyo16}. See \citealp{pueyo18} for a review), direct imaging techniques have been rapidly developing in the past five years. As a result, many objects including both exoplanets (e.g., 51~Eri: \citealp{macintosh15, rajan17}; PDS~70: \citealp{keppler18, muller18, christiaens19}) and circumstellar disks (e.g., HD~106906: \citealp{kalas15, lagrange16}; HD~61005: \citealp{esposito16}; HD~114082: \citealp{wahhaj16}; HIP~73145 [HD~131835]: \citealp{feldt17}; HD~129590: \citealp{matthews17}) have been discovered and/or characterized.

In the post-processing of direct imaging data, the fundamental step is to properly model and remove the star light and speckles (i.e., point spread function, PSF) from the observations. Existing PSF modeling methods have been optimized for unresolved exoplanets where the point spread function is known \citep{marois10, pueyo16}, or for circumstellar disks in space-based observations where the PSF is relatively stable \citep{ren18}. For ground-based observations, polarization differential imaging (PDI) has revealed complex circumstellar structures \citep[e.g.,][]{avenhaus18, monnier19, garufi20}. However, PDI requires linear polarization of the signals, which is typically not present for planets. For disks, such signals can be too faint and thus beyond detection. For the disks detected through PDI, total intensity signals will help better understand the dust properties.

To characterize circumstellar disks in total intensity, post-processing strategies such as reference differential imaging \citep[RDI:][]{smith84}, angular differential imaging \citep[ADI:][]{marois06} and spectral differential imaging \citep[SDI:][]{marios00, sparks02} are adopted. These strategies are accompanied with contaminations from quasi-static and rapidly varying speckles \citep{pueyo18}, and the contaminations are expected be reduced with advanced post-processing methods (e.g., Locally Optimized Combination of Images [LOCI]: \citealp{lafreniere07}; Karhunen-Lo\`eve Image Projection [KLIP]: \citealp{soummer12, amara12}). Nevertheless, existing post-processing methods are limited by over-fitting and self-subtraction issues, and forward modeling is required to recover the surface brightness distribution of the circumstellar objects. However, forward modeling is currently only optimized for unresolved point sources \citep{marois10, pueyo16}. For resolved circumstellar disks, forward modeling needs assumptions on both morphology and flux (e.g., space-based: \citealp{choquet16, choquet17, choquet18}; ground-based: \citealp{follette17, esposito18}), which is model-dependent thus likely unable to capture the minute irregularly-shaped structures (e.g., the disks in \citealp{avenhaus18, monnier19, garufi20}).

To properly separate circumstellar disk signals from observations, \citet{ren18} has demonstrated the applicability of sequential non-negative matrix factorization (sNMF) to space-based observations. For these observations with stable PSFs, sNMF is able to resolve the over-fitting problem for RDI observations after adopting a scaling factor. However, sNMF is still limited by not only unstable PSFs where multiple scaling factors are needed, but also ADI or SDI post-processing where self-subtraction still persists.

In this paper, we aim at extracting thus characterizing circumstellar signals in high contrast imaging observations using a data imputation approach. In multivariate statistical analysis, data imputation is a process which fills missing data based on their relationship with known data \citep[e.g.,][]{johnson2007applied}. For high contrast imaging data, specifically, we first flag the regions that host circumstellar signals as ``missing data'' based on prior knowledge of their location. We then replace the ``missing data'' regions with PSF-only signals (i.e., data imputation). The circumstellar signal then resides in the residual image when the data imputation model is removed from the original image. In this paper, we study the data imputation property of sNMF, i.e., DI-sNMF.

The structure of this paper is as follows: in Section~\ref{sec-method} we describe the DI-sNMF method. In Section~\ref{sec-planet}, we apply it to a simulated planet dataset. In Section~\ref{sec-disk}, we apply it to circumstellar disk simulation and observation. We discuss and summarize the paper in Section~\ref{sec-sum}. We present our mathematical derivations and proofs in \ref{append-nmf-di} and \ref{di:model}, and the auxiliary figures in \ref{append-b}.

\section{Methods}\label{sec-method}
The NMF method was first introduced by \citet{paatero94}, then \citet{lee01} provided multiplicative update rules to guarantee its convergence. Its application to astronomical data started with \citet{blanton07}, where weighting terms were added to take into account of observational uncertainties. \citet{zhu16} then proposed a parallel form in studying spectroscopic data in astronomy to increase computational efficiency. \citet{ren18} demonstrated the applicability of NMF to high contrast imaging with a sequential construction of the components (i.e., sNMF).

\subsection{sNMF for High Contrast Imaging}

In the post-processing of high contrast imaging data, we first have $N_{\rm ref}$ reference images with each image having $N_{\rm pix}$ pixels. For each image, we flatten it to a 1-dimensional row matrix of length $N_{\rm pix}$ by reordering the elements from the original 2-dimensional matrix. Following the definition of symbols in \citet{ren18}, we denote the real-valued $N_{\rm ref}$-by-$N_{\rm pix}$ reference matrix by $R\in\mathbb{R}^{N_{\rm ref}\times N_{\rm pix}}$, which contains $N_{\rm ref}$ flattened reference images. The NMF method performs dimension reduction via approximating $R$ with the product of two matrices (i.e., $R\approx WH$): a coefficient matrix $W\in\mathbb{R}^{N_{\rm ref}\times n}$, and a component matrix $H\in\mathbb{R}^{n\times N_{\rm pix}}$, where $n$ is the number of components. In the component matrix, each row is an NMF component that is expected to be a representation of at least a fraction of the signals in $R$.

\paragraph{Component Construction}The decomposition of $R$ is achieved with the following vectorized weighted multiplicative update rules \citep{zhu16}:
\begin{align}
W^{(k+1)} &= W^{(k)}\circ\frac{(V\circ R)H^{(k)T}}{[V\circ W^{(k)}H^{(k)}]H^{(k)T}}, \label{eq-a1:coef}\\
H^{(k+1)} &= H^{(k)}\circ\frac{W^{(k)T}(V\circ R)}{W^{(k)T}[V\circ W^{(k)}H^{(k)}]}, \label{eq-a2:comp}
\end{align}
where superscript $^{(k)}$ and $^{(k+1)}$ are iteration steps $k$ and $(k+1)$, respectively. Superscript $^T$ denotes matrix transpose. $V\in\mathbb{R}^{N_{\rm ref}\times N_{\rm pix}}$ is a weighting matrix (which is the element-wise inverse of the squared uncertainties of $R$ in \citealp{zhu16}). Symbol $\circ$ and fraction bar $\frac{(\cdots)}{(\cdots)}$ are element-wise multiplication and division of two matrices of the identical dimension, respectively. In the iteration, the initialization values are uniformly drawn from 0 to 1, and the convergence of the above iteration rules is guaranteed when the iteration number increases \citep{zhu16}. Consequently, this initialization guarantees non-negativity for $W$ and $H$. For an intuitive understanding of the component construction update rules, if we ignore the weighting matrix $V$ and the coefficient matrix $W^{(k)}$ on the righthand side of Equation~\eqref{eq-a1:coef}, we have\[W = \frac{R H^T}{H H^T},\] which is a vector projection of $R$ to $H$. Similarly, the intuitive understanding of Equation~\eqref{eq-a2:comp} is the symbolic conjugate of Equation~\eqref{eq-a1:coef}. In other words, the update rules performs least square estimation but with all the matrix elements non-negative \citep{ren18}. Nevertheless, we note that such an approach is not mathematically established, and it is for the purpose of intuitive understanding of NMF update rules only.

To rank the contribution from each component, \citet{ren18} studied the sequential construction of NMF components (i.e., sNMF). Specifically, given $m$ constructed components ($m=1, 2, \cdots, n-1$), the construction of the $(m+1)$-th component is initialized with the converged values of the first $m$ components in the first $m$ rows of $H$, with the $(m+1)$-th row of $H$ still uniformly drawn from $0$ to $1$. 

\paragraph{Target Modeling}For a flattened target image $T\in\mathbb{R}^{1\times N_{\rm pix}}$, it can be modeled by the sNMF components $H$ constructed above. The coefficients are obtained with the following update rule:
\begin{equation}
\omega^{(k+1)} = \omega^{(k)}\circ\frac{(v\circ T)H^T}{[v \circ \omega^{(k)}H]H^T},\label{eq-a3:model}
\end{equation}
where $v\in\mathbb{R}^{1\times N_{\rm pix}}$ is the weighting matrix for $T$, and $\omega\in\mathbb{R}^{1\times n}$ is the coefficient matrix for $T$. This modeling procedure is expected to be linear to the first order \citep{ren18}, i.e., the signal of the circumstellar disk or planet can be linearly separated from the PSF. For an intuitive understanding of the target modeling update rule, we can ignore the $v$ and $\omega^{(k)}$ terms in Equation~\eqref{eq-a3:model} and obtain a least square estimation in the form of \[\omega = \frac{T H^T}{HH^T},\] which is obtaining the coefficients the projection of the target $T$ onto the component basis $H$, and thus it is effectively a special case of Equation~\eqref{eq-a1:coef}. Nevertheless, similarly as for the intuitive understanding of the component construction update rules, we readdress such an approach is not mathematically founded and is for illustration purpose only.

\paragraph{sNMF with ``Missing Data''}When there is ``missing data'' in $R$ or $T$, Equations~\eqref{eq-a1:coef}--\eqref{eq-a3:model} can handle and ignore them during component construction or target modeling. This is achieved by first flagging the missing data with a binary mask, and then ignoring them during component construction and/or target modeling. Specifically, the mask is set to 0 when there is circumstellar signal and 1 otherwise. This binary mask concept is identical to the mask in previous methods in high contrast imaging \citep[e.g.,][]{pueyo12, milli12}. For sNMF, we neglect the contribution from circumstellar signals by treating their corresponding regions as mock ``missing data'' in the sequential construction of the sNMF component basis. We note that this approach ignores the matrix elements, rather than affecting the update rules by replacing these elements with 0. We prove that this element ignoring approach does not bias the sNMF components with non-PSF signals or zero values in Appendix~\ref{di:comp}. We quantify in Appendix~\ref{di:comps-md} that the bias introduced by missing data is negligible when the masked out region is relatively small.

\begin{figure*}[hbt!]
\center
\includegraphics[width=\textwidth]{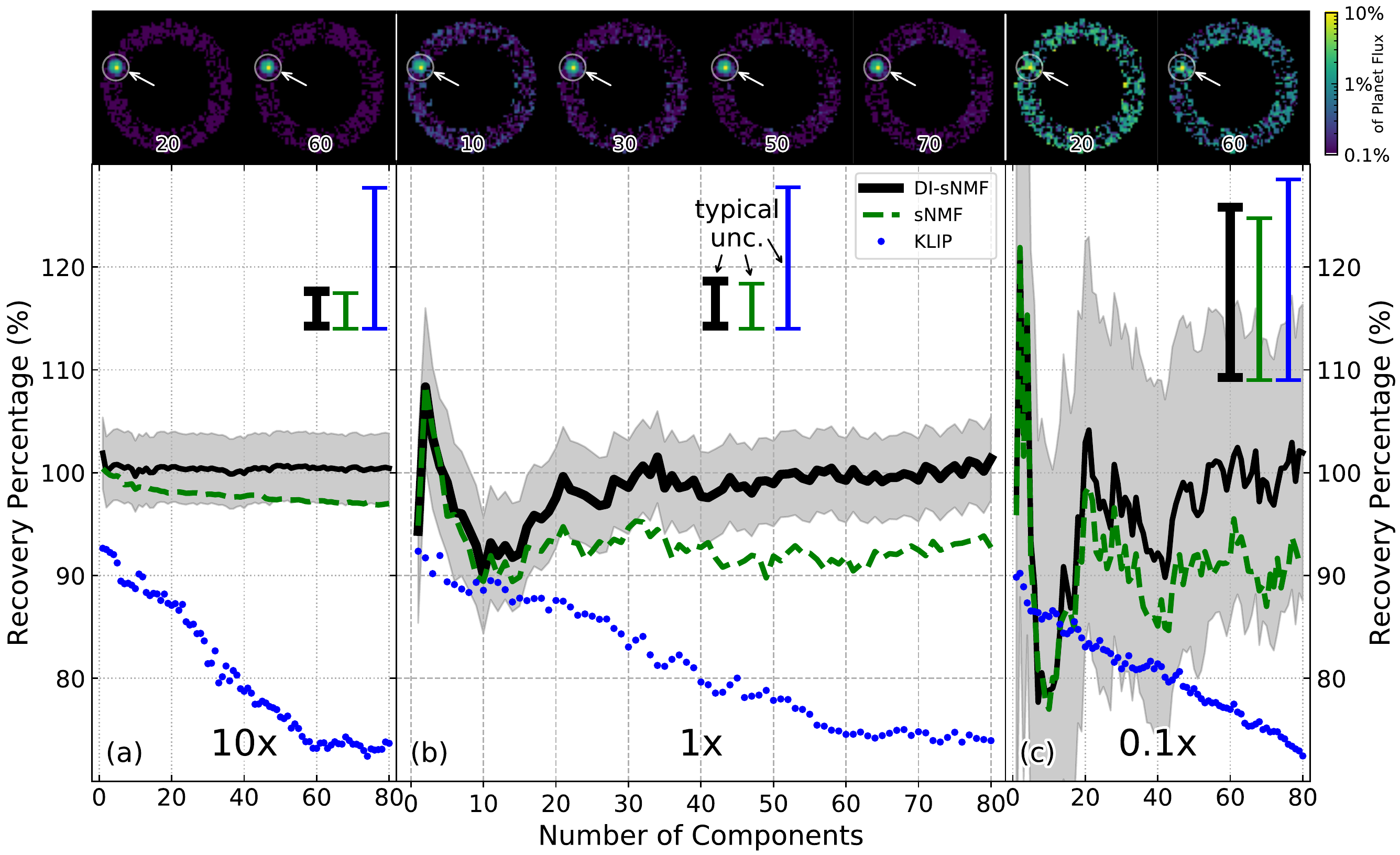}
\caption{Aperture photometry RDI recovery percentages of planet signals injected to \textit{HST}/STIS BAR5 observations of HD~38393. The injected planets are (\textbf{a}) 10${\times}$, (\textbf{b}) 1${\times}$, and (\textbf{c}) 0.1${\times}$ the brightness of the speckles at $r=1\farcs25$. In (\textbf{b}), DI-sNMF is able to recover the injected planet signal at $1\sigma$ levels within ${\sim}5\%$ with decreasing bias when the number of components increases, which outperforms sNMF and KLIP. For the bright planet in (\textbf{a}), DI-sNMF robustly recovers it with small fractional uncertainty. For the faint planet in (\textbf{c}), DI-sNMF is still able to recover the signal albeit with larger fractional uncertainty. Note: (1) The photometry aperture is a $3{\times}3$ pixel square, and we conservatively estimate the uncertainties using the square root of the quadratic sum of the element-wise standard deviation across the $90$ residual images inside the aperture. (2) To reduce the cluttering of the plots, we only present the uncertainties for DI-sNMF in shaded areas, and the typical uncertainties for other reduction results are presented at the top of the corresponding plots. (3) The normalized DI-sNMF images (dimension: $87{\times}87$ pixel or $4\farcs41{\times}4\farcs41$), obtained by dividing the reduction results by the total brightness of the corresponding injected planet, are shown in log scale from $0.1\%$ to $10\%$ of the total brightness of the  planet models. The numbers of components used are presented at the bottom of the images. (4) At the location of the injected point source, the photon noise of the PSF wing is ${\sim}6\%$ per pixel.}
\label{fig:planet-model}
\end{figure*}

\subsection{Data Imputation for ``Missing Data''}\label{sec-di-md}
The data imputation approach imputes the mock ``missing data'' back by modeling an observation with the sNMF components obtained in the previous subsection. Specifically, in target modeling, we obtain the coefficients for a target using Equation~\eqref{eq-a3:model} while ignoring the ``missing data'' region for both the target and the component basis (i.e., the partial target and the partial component basis). We denote the corresponding coefficients by $\omega'$.

We apply $\omega'$ to the whole component basis $H$ and obtain the data imputation model, \begin{equation}
\widetilde{T}=\omega' H\label{eq-DI-sNMF}
\end{equation}
where $\widetilde{T}\in\mathbb{R}^{1\times N_{\rm pix}}$ is the DI-sNMF model of $T$. The missing data region is now imputed with PSF-only signals.

We now establish the mathematical foundation for the data imputation procedure for DI-sNMF. On one hand, we prove in Appendix~\ref{di:model1} that when the non-PSF signals are neglected, they do not bias the coefficients obtained from Equation~\eqref{eq-a3:model}, i.e., $\omega'$. These coefficients are applied to the full sNMF component basis in Equation~\eqref{eq-DI-sNMF} to obtain an empirical model of the ``missing data'' region using PSF-only signals. On the other hand, when the ``missing data'' region is relatively small, these coefficients are expected to have negligible deviation from the coefficients in the scenario where there is only  PSF signals in the entire field of view, see Appendix~\ref{di:coefs} for the proof.

We develop the {\tt nmf\_imaging} package \citep{nmfimaging}\footnote{\url{https://github.com/seawander/nmf_imaging}} for the above steps, i.e., component construction and target modeling with ``missing data'' using binary masks.

\subsection{Circumstellar Signal Extraction}\label{sec-di-md-disks}
When the DI-sNMF model in Equation~\eqref{eq-DI-sNMF} is removed from the original target image, the residual map,
\begin{equation}
\widetilde{D} = T - \widetilde{T},\label{eq-di-disk}
\end{equation}
where $\widetilde{D}\in\mathbb{R}^{1\times N_{\rm pix}}$, is expected to host the circumstellar signals (i.e., non-PSF signals).
\section{Exoplanet Imaging}\label{sec-planet}
\subsection{Data Reduction}\label{sec-planet-process}
We first demonstrate the DI-sNMF method using RDI post-processing of a point source. We use the \textit{Hubble Space Telescope} (\textit{HST})/STIS coronagraphic imaging observations of HD~38393 behind the BAR5 occulter \citep{schneider17, debes19} for this purpose. HD~38393 was visited with 9 orbits during \textit{HST} Cycle 23 (Proposal ID: 14426, PI: J.~Debes), each with a distinct telescope orientation and a $3{\times}3$ sub-pixel dithering pattern. In each dithering position of a visit, a 0.2 s readout is repeated for 10 times. These data have been used in \citet{ren18}, here we only focus on the observations at the central dithering location, i.e., a total of 90 images ($=$ 9 telescope orientations $\times$ 1 dithering position $\times$ 10 readouts). The $90$ images are of dimension $87{\times}87$ pixel ($4\farcs41{\times}4\farcs41$), and we focus only on the pixels with angular separation between $1\farcs0$ and $1\farcs5$. To simulate STIS' instrumental response to an unresolved planet signal in visible light, we obtain the {\tt TinyTim} point source model \citep{krist11}\footnote{\url{http://www.stsci.edu/software/tinytim/}} for a G0 spectral energy distribution.

We simulate RDI post-processing as follows. We first inject a {\tt TinyTim} point source at certain brightness with Poisson noise to the observations of one telescope orientation (i.e., a total of $10$ images). We position the point source at a physical location with radial separation $r=1\farcs25$ and position angle PA $=70^\circ$ (North to East) to avoid occultation of the planet by the occulter or by the diffraction spikes, thus ensuring the maximum coverage of the planet and forming the target exposures. We then construct the KLIP and sNMF component bases using the rest of the $80$ reference images that are planet-free. The $10$ target images are then reduced with KLIP, sNMF, and DI-sNMF. For DI-sNMF, specifically, we use the basis to impute the ``missing data'' region---a $3{\times}3$ pixel region\footnote{The size of this region is identical to region that is used to determine the point source contrast for STIS BAR5 in \citet{debes19}.} where the planet is expected to reside---with PSF-only signals. We repeat this injection subtraction procedure for $9$ times for all of the telescope observations, and calculate the element-wise median and standard deviation for the $90$ PSF-subtracted residual images.

We repeat the above procedure 3 times for a planet at different brightness levels---one that is as bright as the speckles (i.e., in the $3{\times}3$ pixel box, the total planet signal is equal to the total speckle signal), and two that are either $10$ or $0.1$ times of that. For the results obtained from all the reduction methods, we measure the brightness of the point source following \citet{debes19}: we first integrate over the $3{\times}3$ pixel box for the element-wise median of the $90$ PSF-subtracted images, then subtract that value by $9$ times the median of the surrounding pixel values (that are within $2$ and $4$ pixels from the point source center) to remove the background. We calculate a conservative uncertainty\footnote{The uncertainties in this paper are $1\sigma$ unless otherwise specified.} by taking the square root of the quadratic sum of the element-wise standard deviation of the $90$ PSF-subtracted images in the $3{\times}3$ pixel box. We present the recovery fraction results, as well as some recovery fraction images from DI-sNMF, in Figure~\ref{fig:planet-model}.

\subsection{Results}

In Figure~\ref{fig:planet-model}, we observe that DI-sNMF is able to robustly recover the injected planet at various brightness levels. When more DI-sNMF components are used, the recovery is better. Specifically, for a planet that is as bright as the speckles, we notice that DI-sNMF is able to properly recover its flux within $5\%$ at $1\sigma$. We notice that the fractional noise for the point source that is $10{\times}$ the brightness of the speckles is similar but slightly smaller than that of $1{\times}$. This indicates that the variation introduced by PSF subtraction overwhelms the Poisson noise from the injected planet. When the planet is $0.1{\times}$ the brightness of the speckles, both PSF subtraction variation and Poisson noise contribute to the uncertainty, thus resulting to larger fractional uncertainties.

For KLIP, we confirm its expected over-subtraction behavior, i.e., the recovered planet flux decreases when more components are used \citep{soummer12}. In this way, KLIP requires forward modeling to recover the original flux for the planet \citep[i.e., KLIP-FM: ][]{pueyo16}. Forward modeling requires prior knowledge of the 2-dimensional distribution of the unresolved planet signal \citep{pueyo16}, however this information is not required in DI-sNMF. Given that DI-sNMF only requires a 2-dimensional mask to flag the ``missing data'' region, it is therefore another independent planet characterization method with less prior requirement. In addition, DI-sNMF is able to achieve smaller uncertainties than KLIP.  

For sNMF, we observe a less aggressive over-subtraction than KLIP, and sNMF still requires a scaling factor to recover the flux as studied in \citet{ren18}. However, the adoption of a scaling factor is expected to increase the overall brightness distribution in the field of view, thus increasing the overall background noise. For unstable PSF images such as ground-based observations, multiple scaling factors are needed to better recover the point source and its surrounding background, however such approach is currently not well-founded \citep{ren18}. Instead, we study the DI-sNMF method to avoid this problem in a mathematically well-established way.

In this section, we only performed DI-sNMF for the RDI target modeling process. For point source observations with ADI post-processing, we can mimic the RDI procedure by masking out the point sources during both component construction and target modeling (see similar approaches in \citealp{pueyo16} and references therein). For observations with SDI post-processing, we can mimic RDI by spatially scaling the observations at different wavelengths according to their wavelengths and masking out the planet sources. We do not demonstrate the ADI, SDI, or ADI + SDI capabilities of DI-sNMF for point sources since there is no mathematical difference between them and RDI. For the DI-sNMF applicability in the component construction process, we demonstrate it in a more complicated scenario with circumstellar disks in the following section.

\section{Circumstellar Disk Imaging}\label{sec-disk}
\begin{figure*}[hbt!]
\center
\includegraphics[width=\textwidth]{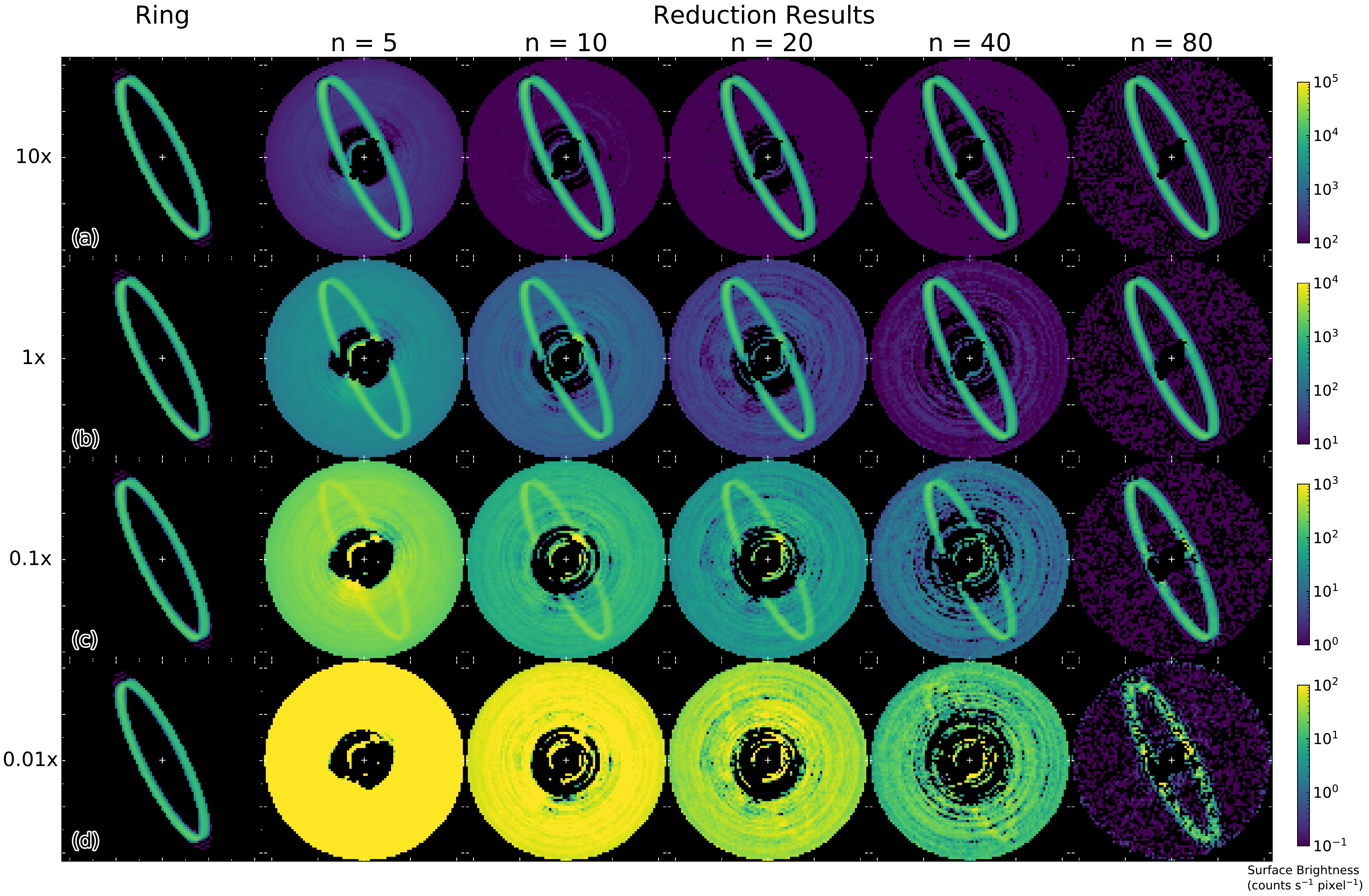}
\caption{Data imputation for simulated ADI post-processing of an isotropic ring (dimension of each image: $87{\times}87$ pixel or $4\farcs41{\times}4\farcs41$). Different rows correspond to the disks that are $10{\times}$, $1{\times}$, $0.1{\times}$, and $0.01{\times}$ the brightness of the PSF wing, respectively. When more components are used the disks are better recovered. When the disk is $0.01{\times}$ the brightness of the PSF, its recovery is marginal. See Figure~\ref{fig:ring-residual} for the corresponding residual maps. Note: each row is at different brightness levels, and each color bar is only for the corresponding row.}
\label{fig:ring-simulation}
\end{figure*}

\begin{figure*}[hbt!]
\center
\includegraphics[width=\textwidth]{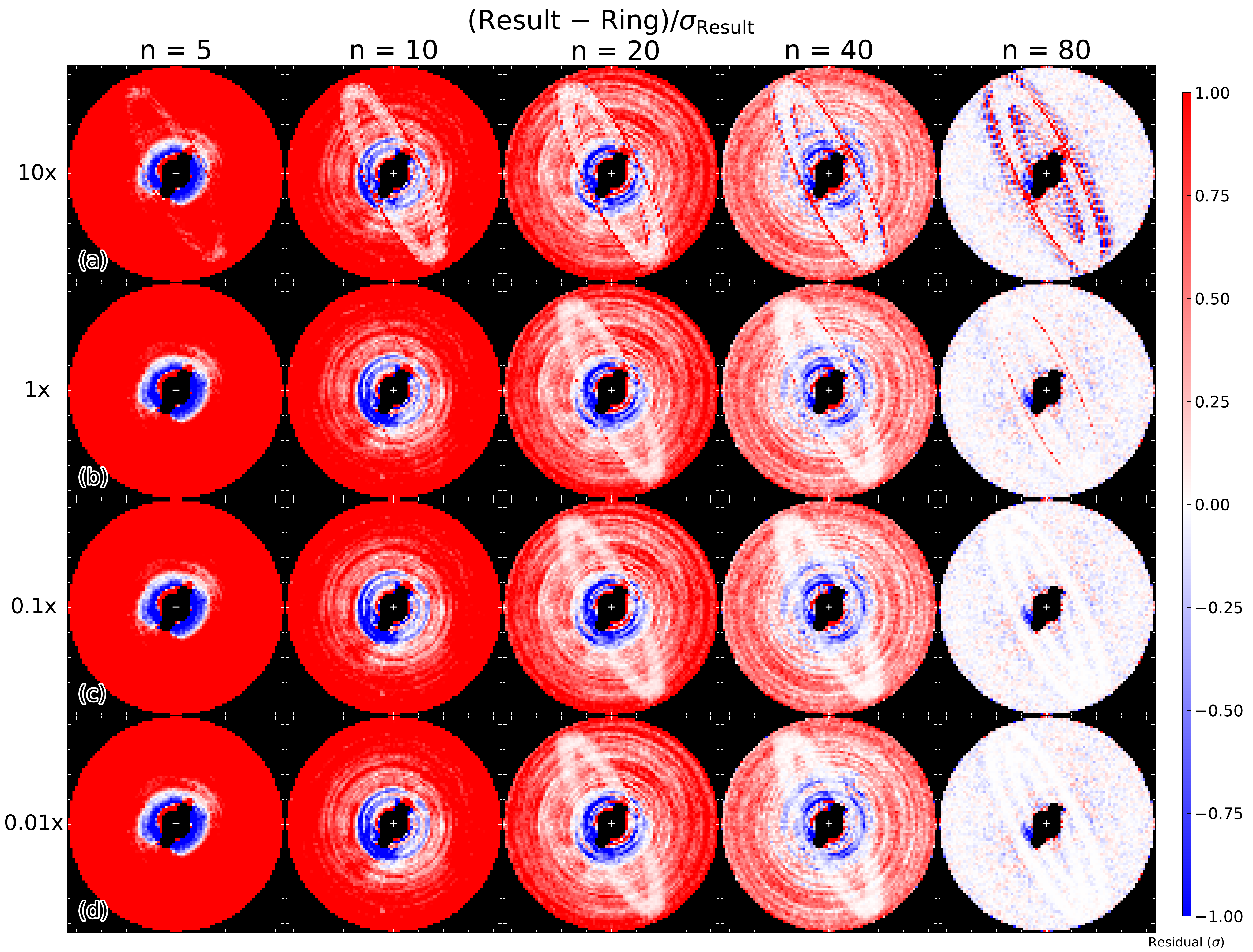}
\caption{Residual maps for Figure~\ref{fig:ring-simulation}. The residual maps are obtained by comparing the median of the DI-sNMF results with the ring models, and divide that by the standard deviation of the DI-sNMF results for different scenarios. In all scenarios, the disk model can be better recovered when more components are used. When all the components are used, the disks are recovered within $0.5\sigma$.}
\label{fig:ring-residual}
\end{figure*}

\begin{figure*}[hbt!]
\center
\includegraphics[width=\textwidth]{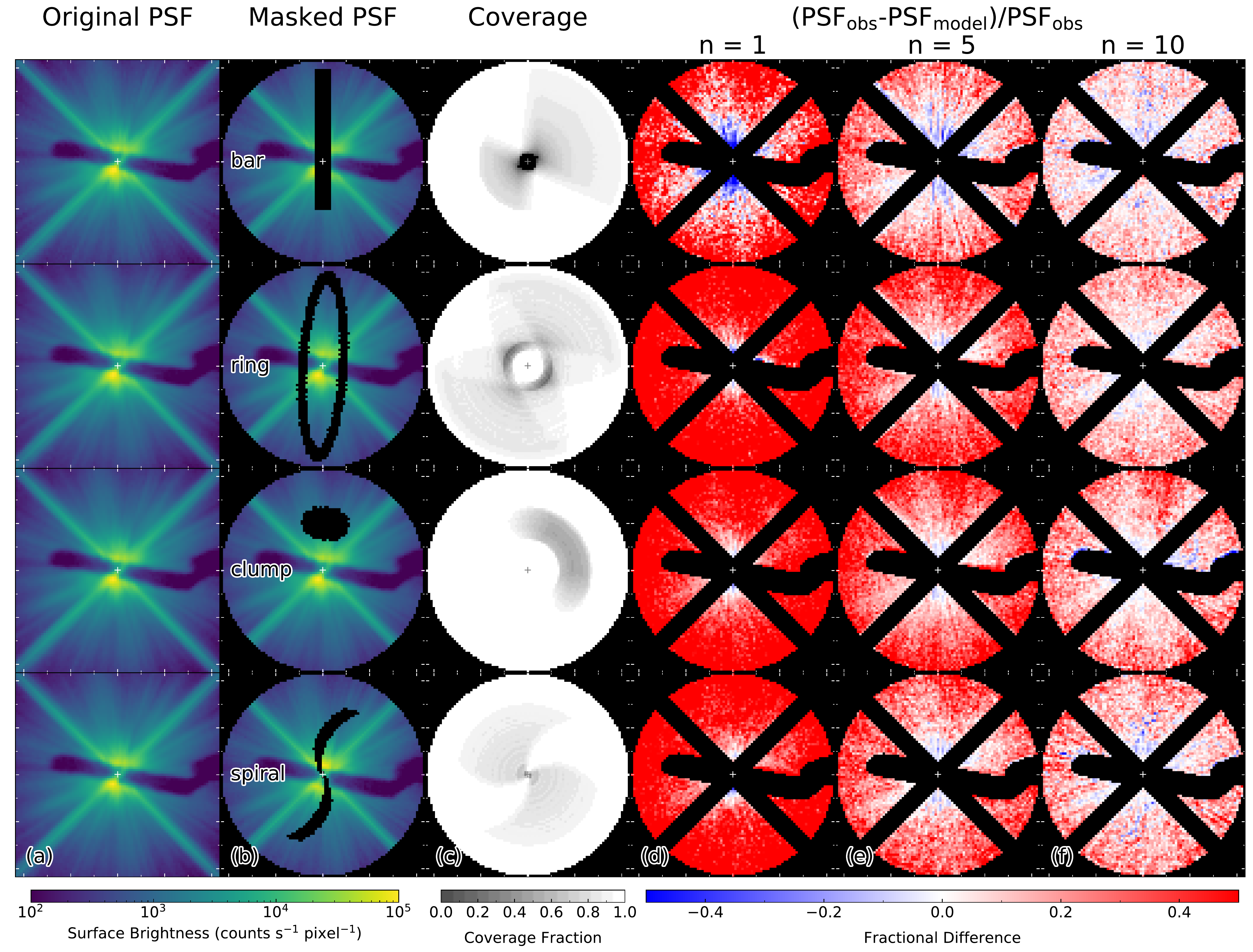}
\caption{Data imputation for simulated ADI post-processing in four scenarios: bar, ring, clump, and spiral. Columns (\textbf{a}) and (\textbf{b}): original and masked exposures. Column (\textbf{c}): coverage fraction of the field of view when the structures are excluded in the simulated ADI sequence. Most of the pixels are covered by ${>}80\%$ of the $81$ images, thus allowing for the PSF signal capture during component construction. Columns (\textbf{d}), (\textbf{e}) and (\textbf{f}): fractional difference between a single frame (i.e., column a) and the DI-sNMF PSF models for the masked frame (i.e., column b) with $n=1, 5,$ and $10$ components, respectively. The recovery quality increases as more components are used. When $10$ components are used, the masked region in column (\textbf{b}) are recovered within ${\sim}10\%$. Note: all of the images here are of dimension $87{\times}87$ pixel ($4\farcs41{\times}4\farcs41$).}
\label{fig:disks-simulation}
\end{figure*}

\subsection{Simulation (ADI)}\label{sec-adi-isodisk}
\subsubsection{Data Reduction}
We now explore the capability of DI-sNMF in terms of frame diversity and mask overlap, which can be translated to a simulated ADI post-processing scenario. We first assemble from the original \textit{HST}/STIS 810 images of HD~38393 in Section~\ref{sec-planet-process}, but instead sample one image from each dithering position at each telescope orientation, totaling $81$ images ($=$ 9 telescope orientations $\times$ 9 dithering position $\times$ 1 readout) to obtain moderate variation of the signals. To simulate observations for ADI post-processing,  we treat these $81$ images as independent PSF realizations that do not depend on telescope orientations. We assign $1\fdg5$ separations among adjacent images, thus obtaining a total field rotation of $120\fdg$ During data reduction, we reduce the impact from STIS' diffraction spikes and the BAR5 occulter using a mask created by \citet{debes17}. For a specific application to a ground-based dataset, see Section~\ref{adi-sdi-data-reduction}.

\paragraph{Simple Ring}To simulate ADI data reduction on mock ``missing data'', we first demonstrate the DI-sNMF method using an isotropic scattering ring generated by the {\tt MCFOST} radiative transfer software \citep{pinte06, pinte09}. We inject the ring to the $81$ images, construct the components by masking out the region containing the ring, and impute the ring region with PSF signals using DI-sNMF. We calculate the element-wise median of the derotated and reduced images to obtain the final result, and calculate the element-wise standard deviation of these images to obtain the noise map. We generate the ring at four brightness levels---the disk are $10{\times}$, $1{\times}$, $0.1{\times}$, and $0.01{\times}$ the brightness of the PSF wing per pixel, respectively. We present the ring models and reduction results in Figure~\ref{fig:ring-simulation}, and the residual maps in Figure~\ref{fig:ring-residual}.

\paragraph{Complex Structures}Being aware of different circumstellar structures, we generalize the ring study and mask out four different regions---bar, ring, clump, and spiral arms. Noticing the fact that the recovery quality of circumstellar signals depends only on the data imputation quality for the PSF signals (see \ref{append-nmf-di}), we do not perform the injection-recovery process as for the planet or the ring model. Instead, we focus on the recovery quality of the PSF signals when the original values are known for a specific frame. For each circumstellar structure, we construct the component basis using the masked PSF signals, then compare the recovered image with the original (i.e., non-masked) image of the frame. We present the results from the ADI reduction with DI-sNMF for these structures in Figure~\ref{fig:disks-simulation}.  

\subsubsection{Results}
For the ring disk model, we notice in Figure~\ref{fig:ring-simulation} that DI-sNMF with ADI can detect the disk from $10{\times}$ down to $0.1{\times}$ the brightness of the PSF wing. When the disk is $0.01{\times}$ the PSF wing, the detection is marginal since it is below the photon noise of ${\sim}0.06{\times}$ the PSF wing. For all brightness levels, more components are likely better recover the ring brightness. In the residual maps in Figure~\ref{fig:ring-residual} , we notice that when more components are used, the detection of the ring is reached within $0.5\sigma$.

Noticing the ring model might be a rare case of well-recoverable geometry, we present the PSF recovery fractions for a single frame under the four different scenarios in Figure~\ref{fig:disks-simulation}. Even though the ideal approach is to perform inject-recovery exercises for these scenarios, we do not implement this approach since we expect the recovery of circumstellar signals to depend only on the recovery of the PSF signals (\ref{append-nmf-di}). When we exclude the circumstellar structures during component construction, most of the field of view is covered by at least $80\%$ of the images (Figure~\ref{fig:disks-simulation}c), which thus allows for the capture of the PSF signals with DI-sNMF. 

Using the DI-sNMF components, we notice that when we increase the number of components used in modeling, the recovered values for the pixels are closer to the observation. When $10$ components are used (Figure~\ref{fig:disks-simulation}f), most of the pixels are recovered with an error of less than ${\sim}5\%$, and the masked regions are recovered with an error of less than ${\sim}10\%$. In other words, a recovery precision of ${\sim}10\%$ for the PSF signals thus allows for the recovery of circumstellar signals that are at least ${\sim}10\%$ the brightness of the PSF wing or speckles.

For the point source simulation in Section~\ref{sec-planet}, we observe that a few components are able to properly recover the flux. However for extended structures, we observe that using too few components can lead to insufficient recovery for the PSF signals, see Figure~\ref{fig:disks-simulation} where using too few components may cause systematic offsets in PSF recovery thus biasing circumstellar structure extraction.

\subsection{Application (ADI + SDI)}\label{adi-sdi-data-reduction}
\subsubsection{Data Reduction}
We demonstrate the ADI + SDI capability of DI-sNMF with the ground-based observations of a debris disk---the HR~4796A system that hosts a narrow disk \citep{mouillet97, augereau99, schneider99, schneider09, schneider18, perrin15, debes08, milli15, milli17, milli19}. We obtained the Gemini/GPI $K1$-band coronagraphic observations of the system on 2015 April 03 (``$K1$-coron''; GS-2015A-Q-27, PI C.~Chen). The GPI $K1$-coron data at $1.89\,\mu$m--$2.20\,\mu$m ($R\approx62$--$70$, with $37$ channels) are taken with $90$s exposure, totaling $48$ exposures with a total field rotation of $78\fdg5$ (a total of $48\times37=1776$ images).

To recover the disk with DI-sNMF, which is marginally visible at far separations but dominated by speckles in the close-in regions in Figure~\ref{fig:hrcoverage}, we assume it with an inclination of $76^\circ$ and a position angle for the major axis of $26^\circ$ from North to East \citep{perrin15, milli17}. To create a mask where the disk is excluded for data imputation, we exclude the region that are between $0\farcs99$ and $1\farcs23$ ($70$ and $87$ pixels. $14.166 \pm 0.007$ mas pixel$^{-1}$: \citealp{derosa15}) from the star in the deprojected plane. For each observation, we rotate the original mask to match the corresponding time-averaged parallactic angle. To make the maximum use of available PSF signal, we spatially scale the $37$ channel images based on their wavelength ratios with that of the 1st channel. For each exposure, we then collapse the $37$ channels to $5$ bins via pixel-wise median combination ($6$ channel per bin, excluding the first $5$ and last $2$ channels) to increase the signal quality. 

After the above procedure, we have a total of $240$ images ($5$ bins each with $48$ exposures). Identical procedures are performed on the masks, and the non-zero elements in the masks are assigned to be 1. See Figure~\ref{fig:hrcoverage} in \ref{append-b} for the raw spatially scaled images before collapsing, and the coverage fraction map after the above disk-exclusion procedure. After the above procedure, most of the field is covered by PSF-only signals. We use the $240$ scaled-and-collapsed images for sequantial NMF component construction, with the $240$ masks excluding the disks at different detector locations in each of the images. In Figure~\ref{fig:hrcomp} in \ref{append-b} , we present component examples that are constructed in this way---no disk signal is captured in the components as expected from the mathematical derivation in Appendix~\ref{di:comp}.

Using the above components that cover most of the field of view (that includes the region where the disk resides), we model the disk-free regions with DI-sNMF. For each image, we flag the region where the disk resides as ``missing data'' using the corresponding binary mask created above, then obtain the DI-sNMF coefficients for the disk-excluded regions. The coefficients are then used to impute the disk region with the constructed sNMF components. We then obtain the residual image by subtracting the DI-sNMF model from the original target image. The residual images are then spatially scaled back to their original size on detector, derotated to north up and east left. We obtain the final result by taking the element-wise median of these images, and the noise map using the element-wise standard deviation of these images. The corresponding SNR map is then the ratio between the final result and the noise map.

\subsubsection{Results}\label{sec-disk-result}
\begin{figure}[hbt!]
\center
\includegraphics[width=.5\textwidth]{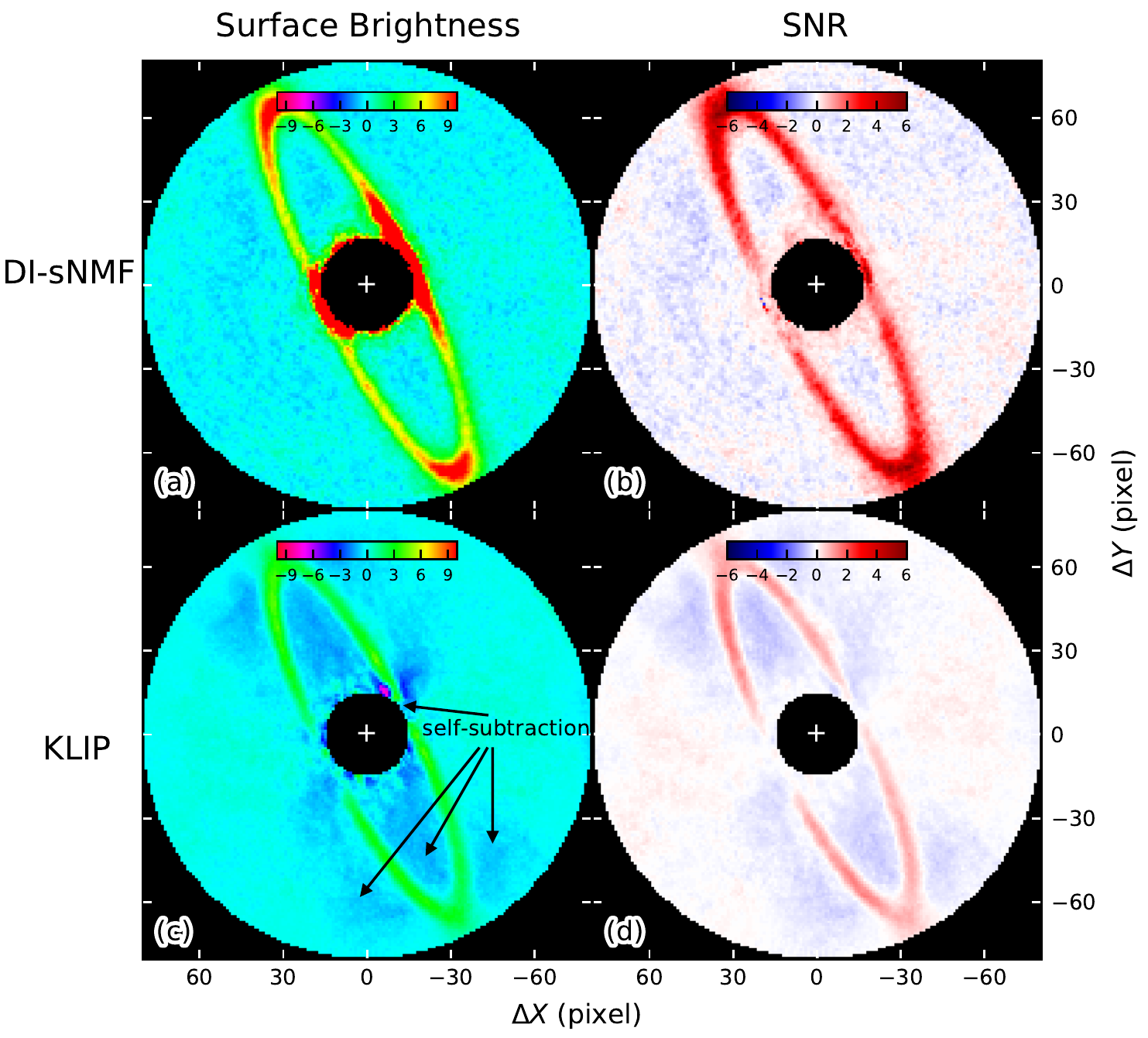}
\caption{Reduction results of the GPI $K1$-coron observations of HR~4796A from DI-sNMF and KLIP. ({\bf a}) Disk image at the central wavelength ($2.05\,\mu$m) obtained from DI-sNMF. ({\bf b}) SNR map for the disk image, obtained by dividing the image by the element-wise standard deviation of the composing $48$ individual images. ({\bf c}) and ({\bf d}) are the corresponding KLIP results, with the negative regions around the disk demonstrating the self-subtraction effect. In comparison, the DI-sNMF result not only has no noticeable negative regions around the disk, which demonstrates that it has no self-subtraction effect, but also has higher SNRs for the disk.}
\label{fig:result-disk}
\end{figure}

\paragraph{Image} Using the above approach, we are able to obtain the disk image at $5$ different bins (i.e., $5$ wavelengths). We present the reduction result\footnote{To reduce confusions from image spatial scaling, the images are all presented in detector counts per pixel in this paper unless otherwise specified.} using $10$ components\footnote{The final disk image does not have significant variation when more than $5$ components are used.} and the corresponding signal to noise ratio (SNR) map for the central wavelength ($2.05\,\mu$m) in Figure~\ref{fig:result-disk}. 

We observe the ring structure with SNR $>5$, and the disk image does not have self-subtraction effects as expected for DI-sNMF. For comparison, we reduce the data using KLIP ADI with 10 components, obtain and present the final image and SNR map in the identical scale as DI-sNMF. In the reduction results, there is noticeable self-subtraction effects in the KLIP result in the forms of negative halos around the disk and reduced surface brightness for the disk, however these effects are not detectable in the DI-sNMF result. What is more, DI-sNMF is able to recover the minor axis of the ring.

We expect that DI-sNMF is able to properly recover the HR~4796A disk in two aspects. On one hand, there are no detectable self-subtraction effects in the DI-sNMF result in Figure~\ref{fig:result-disk}a, indicating that self-subtraction is not impacting the result. On the other hand, the SNRs for the background regions (i.e., non-disk regions) are well-behaved with values of ${\sim}{\pm}1$ in Figure~\ref{fig:result-disk}b, i.e.,
\begin{equation}
\frac{\widetilde{D_i}}{\sigma_i} = \pm 1 + o(1),\label{snr1reduced}
\end{equation}
where $i$ is the pixel index in the non-disk regions\footnote{This definition of $i$ is only applicable within Section~\ref{sec-disk-result}. In other Sections, $i$ refers to any pixel inside the field of view.}, $\sigma_i$ is the corresponding standard deviation, and $o(\cdot)$ is the little $o$ notation (i.e., $|o(x)|\ll |x|$). For independent and normally distributed noise (usually assumed for the PSF subtraction residuals), ${D_i} \sim \mathcal{N}(\mu_i, \sigma_i)$,
where $\mu_i$ is the corresponding expectation. In this way, $\left(\frac{D_i-\mu_i}{\sigma_i}\right)^2$ follows a chi-squared distribution with $1$ degree of freedom (i.e., $\chi^2_1$), 
whose corresponding expectation is $1$. Assuming ideal PSF subtraction (i.e., $\mu_i = 0$), we have the expectation of $\left(\frac{D_i}{\sigma_i}\right)^2$ to be 1. This value is consistent with the squared version of Equation~\eqref{snr1reduced} to the first order, and thus we expect that in our regime of calculating the noise map, the PSF-only signals in the background regions are consistent with ideal PSF reconstruction. Consequently, we expect that the PSF-only signals in the disk-hosting regions are also properly modeled and subtracted, since the constructed PSFs do not have significant deviations between non-masked and masked scenarios when the masked out regions are relatively small (Appendix~\ref{di:coefs}). Nevertheless, we estimate that the deviation has an upper limit of ${\sim}10\%$, since we masked out ${\sim}15\%$ of the field of view for the HR~4696A observations, see the notes in Appendix~\ref{di:coefs} for more specific discussion.

\begin{figure}[hbt!]
\center
\includegraphics[width=.5\textwidth]{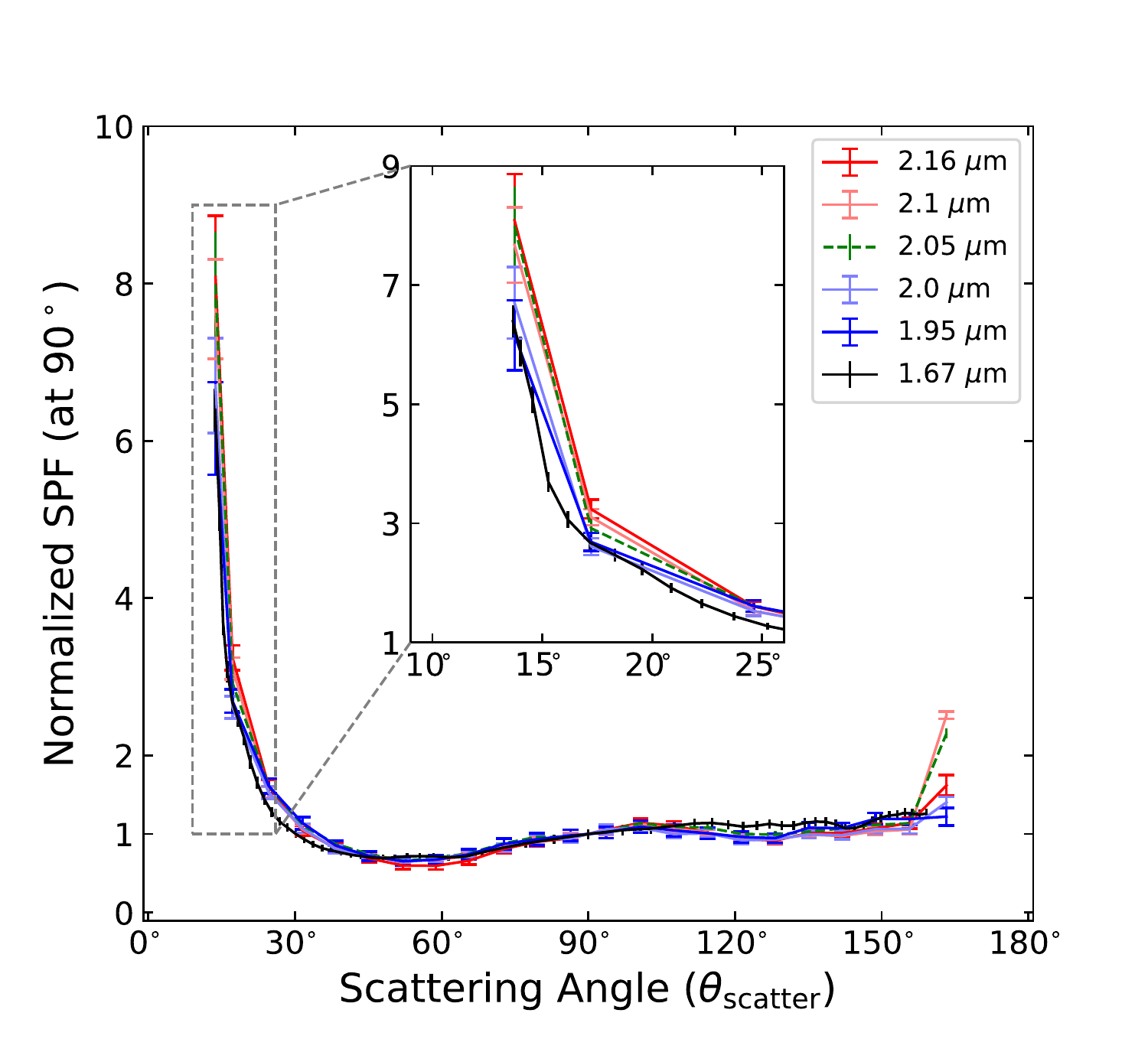}
\caption{Normalized SPFs of the HR~4796A disk in $K1$ observation with DI-sNMF. The $1.67\,\mu$m data (black) is the averaged SPHERE $H2$ SPF in \citet{milli17}. In the probed scattering angles, the inset image suggests that the dust are more forward scattering as wavelength increases.}
\label{fig:spf}
\end{figure}

\paragraph{Scattering Phase Function} Based on the photometry recovery results in Section~\ref{sec-planet}, we do not focus on the brightness of the disk to reduce redundancy. Instead, we extract the surface brightness distribution as a function of scattering angle (i.e., the scattering phase function, SPF: \citealp{hansen74}. See \citealp{hughes18} for a recent collection of SPFs for debris disks). Assuming the disk is circular and there is no offset between the disk center and the star, we obtain the original SPF curves using the equation in \citet{ren19disk}. We then normalize the curves at $90^\circ\pm10^\circ$, and divide them by the original SPF of an isotropic scattering disk created by {\tt MCFOST} to correct for limb-brightening and convolution effects \citep{milli17}. We present the normalized SPFs in Figure~\ref{fig:spf}. 

The $K1$ SPFs are generally consistent with the SPF in the $H2$ band observed using the SPHERE instrument \citep[$1.67\,\mu$m: ][]{milli17}. Nevertheless, for the probed scattering angles, we observe a tentative trend that the SPF is more forward scattering as the wavelength increases. Although the differences are less than $2\sigma$, this trend is expected for the dust in the geometrical scattering regime \citep[e.g., Figure~11 in][]{schroeder14}.

We expect the above trend to be not biased by the symmetric scattering in the most forward and most backward scattering regions (see \citealp{milli17} for the corresponding discussion), since it does not change when we proportionally remove the most backward scattering signals from the most forward scattering regions. In addition, we argue that the halo structure in HR~4796A \citep{schneider18} is likely not biasing the measurements, since it is ${\sim}10\%$ the brightness than the ring, and the $5$ image are reduced in an identical way thus the halo creates a systematic offset for all of them. What is more, the difference of the most forward scattering part in the $5$ SPFs is larger than $10\%$, making it unlikely for the deviation to be originated from the halo. 

To further establish the above SPF trend as a function of wavelength, the comparison between the observations at multiple bands (e.g., $H$, $J$, $K1$, $K2$) is preferred \citep[e.g.,][]{chen19prep}. We do not perform a multi-band experiment since the flux calibration between the different bands need to be properly treated, and more sky rotation are needed to properly recover the SPFs at the other bands \citep{chen19prep}.

\section{Discussion and Summary}\label{sec-sum}
In this paper, we present a new data reduction method, DI-sNMF, for high contrast imaging obervaitons. It is a data imputation approach in both the component construction and the target modeling processes in the extraction of circumstellar objects (exoplanets, disks, etc.). Using one specific property of non-negative matrix factorization, that a fraction of signals can be masked out during both component construction and target modeling, we are able to attribute PSF-only signals to the region where circumstellar objects reside. 

\subsection{Exoplanets}
We demonstrate in this paper that planet signals can be properly recovered in an RDI approach. We construct the sNMF components from full field observations of PSF-only signals, and apply the data imputation approach to the target modeling procedure by masking out the planet and recover the PSF signal at the location of the planet. In our simulation in Section~\ref{sec-planet}, we are able to demonstrate that DI-sNMF is able to not only properly recover the injected flux of the planet, but also obtain smaller uncertainties than KLIP. In this way, DI-sNMF is another independent method to recover planets in addition to the KLIP forward modeling method \citep[KLIP-FM:][]{pueyo16}. Nevertheless, DI-sNMF has less assumptions than KLIP-FM: it only requires a 2-dimensional binary mask, rather than a 3-dimensional distribution of light (i.e., a PSF model for a point source). 

Similarly as for KLIP-FM, a forward model matched filter \citep{ruffio17} can be applied to the entire field of view to discover and characterize planets for DI-sNMF. In addition, although we do not demonstrate the ADI + SDI capability of DI-sNMF for point sources, rotating and spatially scaling the 2-dimensional binary mask will transform the ADI + SDI problem to a classical RDI problem \citep{pueyo16}. Nevertheless, we demonstrate the ADI + SDI scenario for a more complicated problem---circumstellar disks---to show that the DI-sNMF approach is able to achieve the goal of properly recovering the signals, and a point source is just a subset of the more complex circumstellar disk geometry.

\subsection{Circumstellar Disks}
We first demonstrate in this paper that a simple disk model (i.e., an isotropic scattering disk) can be properly recovered in ADI post-processing using DI-sNMF. We then generalize this study, and demonstrate that PSF signals can be properly recovered in various circumstellar structures using simulated ADI approaches. In comparison with point source extraction, more components are needed to better recover the PSF signals.

As an application to disk observations, we perform a precise measurement of the SPF trend for our ADI + SDI post-processing of the debris disk surrounding HR~4796A, finding a tentative trend that the SPFs are more forward scattering as the observation wavelength increases within the $K1$ band. Although a small region on the disk that is closest to the star is not covered due to the limited field rotation (Figure~\ref{fig:hrcoverage}d), the final DI-sNMF image is still able to cover the entire field of view, since this region is then fully covered when we combine the derotated reduced images. In comparison with \citet{chen19prep} where no clear trend of the SPFs is observed at different bands using KLIP ADI forward modeling, we argue that it is the limitation of the KLIP ADI forward modeling method: the minor axis (i.e., the most forward scattering region for the disk surrounding HR~4796A) region suffers from self-subtraction with KLIP ADI (e.g., Figure~\ref{fig:result-disk}c), and forward modeling is unable to probe these values and thus reduce their uncertainties. Therefore the KLIP ADI forward modeling process extrapolates the SPFs for these self-subtraction regions, and such extrapolation depends on the input SPF models which might not be representative of the HR~4796A disk's SPFs. In addition, the uncertainties obtained from forward modeling depends on the likelihood function, however the correlated noises have to be take into account \citep[e.g.,][]{wolff17}. Nevertheless, the SPFs extracted from the DI-sNMF images are empirical measurements that have least dependency on such limitations.

To fully eliminate the not covered region for data imputation in ADI + SDI post-processing, the addition of PSF-only exposures (e.g., \textit{HST}/NICMOS: \citealp{choquet14}; \textit{HST}/STIS: \citealp{ren17}; \textit{Keck}/NIRC2: \citealp{ruane19}) is needed, and a careful selection of them \citep[e.g.,][]{ruane19} is expected to increase the computational efficiency of DI-sNMF. This joint ADI + SDI + RDI approach is expected to recover the disk image with theoretically negligible alteration of the circumstellar signals when the data imputation region is relatively small (\ref{di:model}). Although we demonstrate the DI-sNMF capability for circumstellar structures that have well-defined geometry, we expect that more complex disk structures could be properly recovered---especially for protoplanetary disks which are currently only properly recovered mainly through PDI (e.g., spiral arm systems: \citealp{benisty15, stolker16, monnier19}).

Another promising way to eliminate the not covered region is the iterative approach presented in \citet{pairet18}, which sequentially captures the non-PSF structures since the PSF does not rotate with respect to the detector as the telescope rotates. The combination of sNMF with this approach is likely able to remove the not covered regions in the ADI + SDI approach. However such approach is still not able to handle azimuthally symmetric structures since they do not have distinguishable relative motion from the PSF. In this case, using the RDI references is the unique advantage of DI-sNMF, and the face-on structures are expected to be recovered with minor alteration of their signals.

\subsection{Limitation and Generalization}
In this paper, we only masked out ${\lesssim}15\%$ of the entire field of view to ensure that the DI-sNMF models do not have significant deviation from the true PSF-only signals. In Appendix~\ref{di:coefs} we have derived that the area fraction of the masked region has a second order effect on the deviation. Mathematically speaking, assuming the variance contribution from all the pixels in the field of view are equal, they will have a total deviation of less than ${\sim}2\%$ ($\approx0.15{\times}0.15$). When more areas are masked out (i.e., ${\gtrsim}50\%$), the DI-sNMF method may lead to significant deviations (i.e., ${\gtrsim}25\%$)---we do not further investigate this effect in this paper. However, under this scenario, a possible solution is to increase the field of view to reduce the corresponding impact (for example, the clump example in Figure~\ref{fig:disks-simulation}).

We note that sufficient field rotation is needed for DI-sNMF in ADI post-processing. To construct the sNMF components for the PSF-only signals, the regions hosting non-PSF signals have to be flagged by the binary masks. In addition, the PSF signals within the masked regions should not have variation timescale that is faster than the exposure time of each image, otherwise the DI-sNMF components will not be able to represent the PSF signals and fall into the regime of inadequate data imputation. If these two conditions are not met, a possible solution is the sequential capture of rotating (or non-rotating) signals presented in \citet{pairet18}. In addition, known PSF exposures from available surveys \citep[e.g.,][]{choquet14, ren17, ruane19} will help to loosen these requirements. Despite these limitations, we expect that the most promising application of DI-sNMF is to the thin edge-on disk systems where the required field rotation is smaller.\\

In summary, we have demonstrated that the DI-sNMF approach is expected to properly recover the signals for high contrast imaging observations. It not only is another method for planet characterization, but also opens a new gate to characterize circumstellar disks for integral field spectroscopic observations, and even obtaining their wavelength-dependent image \citep[i.e., reflectance spectra, e.g.,][]{debes08, bhowmik19}. More general applications of DI-sNMF are to the scenarios where the separation of signals is needed---for example, extracting the spectra for the substellar companions in coronagraphic long-slit spectroscopic observations \citep[e.g.,][]{vigan08, hinkley15}.

\facilities{\textit{HST} (STIS), {Gemini:South} (Gemini Planet Imager)}

\software{{\tt MCFOST} \citep{pinte06, pinte09}, {\tt nmf\_imaging} \citep{nmfimaging}, {\tt TinyTim} \citep{krist11}}

\acknowledgments

We appreciate the suggestions in both mathematical expression and main text from the anonymous referee that improved this paper. B.R.~thanks Nicolas Charon and Bingxiao Xu for useful discussions, and Christophe Pinte for sharing the {\tt MCFOST} software and commenting on this paper. Based on observations made with the NASA/ESA \textit{Hubble Space Telescope}, obtained from the data archive at the Space Telescope Science Institute. STScI is operated by the Association of Universities for Research in Astronomy, Inc. under NASA contract NAS 5-26555. Based on observations obtained at the {Gemini Observatory}, which is operated by the Association of Universities for Research in Astronomy, Inc., under a cooperative agreement with the NSF on behalf of the Gemini partnership: the National Science Foundation (United States), National Research Council (Canada), CONICYT (Chile), Ministerio de Ciencia, Tecnolog\'{i}a e Innovaci\'{o}n Productiva (Argentina), Minist\'{e}rio da Ci\^{e}ncia, Tecnologia e Inova\c{c}\~{a}o (Brazil), and Korea Astronomy and Space Science Institute (Republic of Korea). This research project (or part of this research project) was conducted using computational resources (and/or scientific computing services) at the Maryland Advanced Research Computing Center (MARCC: \url{https://www.marcc.jhu.edu}). This work was in part funded by contract 61362448-122362 from WFIRST  NASA project office via a subcontract through Stanford.

\appendix
  
\section{sNMF Component Construction}\label{append-nmf-di}

In this Appendix, we convert the component construction problem for both the ADI and SDI post-processing scenarios to a ``missing data''\footnote{Note: Unless otherwise specified, the ``missing data'' in this paper refers to the regions that are expected to contain circumstellar signals.} problem in statistics. This is achieved by assigning zero weight to the known regions containing circumstellar signals (i.e., circumstellar disks or exoplanets).

\subsection{Component Construction with Missing Data}\label{di:comp}
In an ADI post-processing setup, the sky is rotated relatively to the detector frame on the telescope. In the detector frame, the non-rotating non-circumstellar signals are quasi-static; while the circumstellar signals are rotated around the center of the frame with the sky. Given the position and morphology of these circumstellar signals, we minimize their contribution in the component construction procedure for the {\it non}-circumstellar signals in Equations~\eqref{eq-a1:coef} and \eqref{eq-a2:comp}, and this is achieved by assigning zero weight to these positions in the weighting matrix $V$. 

The assignment of zero weight for positions hosting circumstellar signals thus converts the ADI problem to a missing data problem in statistics, and the data imputation procedure will be able to fill these regions with PSF-only signals. In this subsection, we prove that the zero-weighted ``missing data'' has no contribution to the NMF component construction procedure.

{\bf Lemma 1} \textit{(Components)}: Circumstellar signals do not influence the sNMF component basis of PSF signals when the circumstellar signals are treated as missing data.

{\bf Proof}: We prove this Lemma by way of induction. Let component matrix $H = [H_1^T, H_2^T, \cdots, H_n^T]^T$, where $H_i\in\mathbb{R}^{1\times N_{\rm pix}}$ ($i = 1, 2, \cdots, n)$ is the $i$-th NMF component.

a) When $n = 1$, we have row vector $H = H_1\in\mathbb{R}^{1\times N_{\rm pix}}$ and column vector $W = W_1 \in \mathbb{R}^{N_{\rm ref}\times 1}$, and the iteration rule in Equation~\eqref{eq-a2:comp}, when we ignore the $V$ matrix as in \citet{ren18}\footnote{The proof with $V$ is in principle achieved with the substitutions in Appendix A3 of \citet{blanton07}.}, is rewritten as
\begin{equation}
H_1^{(k+1)} = H_1^{(k)}\circ\frac{W_1^{(k)T} R}{W_1^{(k)T}W_1^{(k)}H_1^{(k)}}, \label{eq-a4:first-comp}
\end{equation}
where the $j$-th element in $H_1$ ($j = 1, 2, \cdots, N_{\rm pix}$) has its iteration rule:
\begin{equation}
H_{1j}^{(k+1)} = H_{1j}^{(k)}\frac{\sum_{i=1}^{N_{\rm ref}} W_{1i}^{(k)} R_{ij}}{\sum_{i=1}^{N_{\rm ref}} W_{1i}^{(k)} \left(W_1^{(k)}H_1^{(k)}\right)_{ij}\ }.\label{eq-a5:elementH}
\end{equation}

The numerator  in Equation~\eqref{eq-a5:elementH} is a real-valued number, since it is the product of a row vector $W_1^{(k)T}$ and the $j$-th column of $R$ (i.e., $R_{(\cdot)j}$). To illustrate the effect of the weighting matrix, especially on its ability of excluding circumstellar signals, we introduce a binary 0-1 indicator matrix of the same dimension as $R$ to represent the exclusion of the circumstellar signals in the weighting matrix, i.e., $\mathbbm{1} \in \mathbb{R}^{N_{\rm ref} \times N_{\rm pix}}$, and for each matrix element indexed $ij$ ($i=1,2, \cdots, N_{\rm ref}$; $j=1,2, \cdots, N_{\rm pix}$), $\mathbbm{1}_{ij} = 0$ means it contains both non-circumstellar and circumstellar signals and $\mathbbm{1}_{ij}=1$ means the it contains only non-circumstellar signals (i.e., PSF signal only). With this representation, we have $R\equiv R\circ \mathbbm{1}$. In this way, the numerator can be rewritten as 
\begin{align}
\sum_{i=1}^{N_{\rm ref}} W_{1i}^{(k)} R_{ij} &\equiv \sum_{i=1}^{N_{\rm ref}} W_{1i}^{(k)} R_{ij} \mathbbm{1}_{ij}\label{eq-a6:indicator-help}\\
	&= \sum_{i=1, \mathbbm{1}_{ij} = 1}^{N_{\rm ref}} W_{1i}^{(k)} R_{ij}\mathbbm{1}_{ij}  + \sum_{i=1, \mathbbm{1}_{ij} = 0}^{N_{\rm ref}} W_{1i}^{(k)} R_{ij} \mathbbm{1}_{ij}\label{eq-a7:two-terms}\\
	&= \sum_{i=1, \mathbbm{1}_{ij} = 1}^{N_{\rm ref}} W_{1i}^{(k)} R_{ij}, \label{eq-a8:individual}
\end{align}
where the second term in Equation~\eqref{eq-a7:two-terms} is $0$, i.e., the zero-valued elements in $R_{(\cdot)j}$ (i.e., when $\mathbbm{1}_{ij} = 0$) do not influence the numerator, and consequently they do not have influence on the iteration of $H_{1j}$. Given that the zero-valued elements in $\mathbbm{1}_{(\cdot) j}$  corresponds to the missing data in $R_{(\cdot) j}$, we have the missing data in $R_{(\cdot)j}$ do not have influence on the iteration of $H_{1j}$. 

Combining all the columns in $R$, the zero-weighted values therefore do not have impact on $H_1$.

b) For $n > 1$, assuming the Lemma holds for the $m$-th component ($m = 1, 2, \cdots, n-1$), then for the $(m+1)$-th component, the sequential construction of the NMF components follows \citep[adapted from Appendix C in][]{ren18}:
\begin{equation}
H_{m+1}^{(k+1)}  = H_{m+1}^{(k)} \circ \frac{W_{m+1}^{(k)T}R}{W_{m+1}^{(k)T}W_{m'}^{(k)}H_{m'}^{(k)} + W_{m+1}^{(k)T}W_{m+1}^{(k)}H_{m+1}^{(k)}},\label{eq-a6:newElementH}
\end{equation}
where $W_{m'}=[W_1, W_2, \cdots, W_m]\in\mathbb{R}^{N_{\rm ref}\times m}$ and $H_{m'}=[H_1^T, H_2^T, \cdots, H_m^T]^T\in\mathbb{R}^{m\times N_{\rm pix}}$ are the first $m$ constructed coefficient and component matrices, respectively.

The numerator  in Equation~\eqref{eq-a6:newElementH} has the same form as that in Equation~\eqref{eq-a5:elementH}, therefore following the identical argument as for $H_1$ from Equation~\eqref{eq-a4:first-comp} to Equation~\eqref{eq-a8:individual}, then if the Lemma holds for $H_m$, the zero-weighted elements in $R$ do not have impact on $H_{m+1}$.

c) Combining the arguments in a) and b), we have proven that if the components are constructed sequentially, the zero-weighted elements in the reference matrix $R$ do not have impact on the component matrix $H$. \QED

In addition, the circumstellar signal--excluded observations in an observation sequence can be added with their references (i.e., the RDI references) to form a new set of PSF-only observations. Following the identical arguments as above, the sNMF component basis for this set contains only PSF signals.

\subsection{Component Variation Induced by Missing Data}\label{di:comps-md}
In the previous subsection, we proved that the sNMF components are not influenced by the circumstellar signals when they are treated as missing data. However, that does not guarantee that missing data does not influence the component basis. In this subsection, we estimate the sNMF basis difference between two scenarios: one that is constructed when there is no missing data, the other that is constructed with missing data. Specifically, we assume the missing data region of the latter is filled with PSF-only signals in the former scenario.

{\bf Theorem 1} \textit{(Components from Missing Data)}: When the signal contribution from the masked region is relatively small with respect to the entire field of view, the sNMF component basis does not have significant variation between the non-masked and masked scenarios.

{\bf Proof}: Without loss of generality, we ignore the $V$ matrix as in Appendix~\ref{di:comp}, and focus on the first column in the coefficient matrix $W$ and the first row in the coefficient matrix $H$ (i.e., $W_1$ and $H_1$, respectively) when $n = 1$.

First, the $i$-th element of $W_1$ in Equation~\eqref{eq-a1:coef} has a form of
\begin{align*}
W_{1i}^{(k+1)} &= W_{1i}^{(k)}\circ\frac{\sum_{j=1}^{N_{\rm pix}}R_{ij}H_{1j}^{(k)}}{W_{1i}^{(k)}\left(\sum_{j=1}^{N_{\rm pix}}H_{1j}^{(k)2}\right)}\\
			&= \frac{\sum_{j=1}^{N_{\rm pix}}R_{ij}H_{1j}^{(k)}}{\sum_{j=1}^{N_{\rm pix}}H_{1j}^{(k)2}}.\numberthis
\end{align*}
When we denote the corresponding coefficient matrix generated by missing data with $W'$, we have
\begin{align*}
W_{1i}'^{(k+1)} &= \frac{\sum_{j=1, \mathbbm{1}_{ij} = 1}^{N_{\rm pix}}R_{ij}H_{1j}^{(k)}}{\sum_{j=1}^{N_{\rm pix}}H_{1j}^{(k)2}}\\
			&= W_{1i}^{(k+1)}-\frac{\sum_{j=1, \mathbbm{1}_{ij} = 0}^{N_{\rm pix}}R_{ij}H_{1j}^{(k)}}{\sum_{j=1}^{N_{\rm pix}}H_{1j}^{(k)2}}.\label{eq-a8-w1i}\numberthis
\end{align*}
Noticing $W_{1i}$ is applied to $H_1$ to approximate the $i$-th row of $R$, i.e., the $i$-th reference image, when the missing data region in the reference image $R_{i(\cdot)}$ is relatively small and does not have significant variation across different images, i.e., 
\begin{equation}
\epsilon_{{\rm image}} \equiv \frac{\sum_{j=1, \mathbbm{1}_{ij} = 0}^{N_{\rm pix}}1}{N_{\rm pix}} \ll 1,  \label{assump-a2-1}
\end{equation}
we can rewrite Equation~\eqref{eq-a8-w1i} as
\begin{align*}
W_{1i}'^{(k+1)} &=  W_{1i}^{(k+1)} - o(W_{1i}^{(k+1)}) \\
			&= W_{1i}^{(k+1)} [1-o(1)], \label{eq-a9-w1i}\numberthis
\end{align*}
where $o(\cdot)$ is the little $o$ notation, i.e., $|o(x)|\ll |x|$. 

Second, the $j$-th element of $H_1$ in Equation~\eqref{eq-a2:comp}, similarly as for $W_{1i}$, has a form of
\begin{align*}
H_{1j}^{(k+1)} &= H_{1j}^{(k)} \circ \frac{\sum_{i=1}^{N_{\rm ref}}W_{1i}^{(k)}R_{ij}}{\left(\sum_{i=1}^{N_{\rm ref}}W_{1i}^{(k)2}\right)H_{1j}^{(k)}}\\
			&= \frac{\sum_{i=1}^{N_{\rm ref}}W_{1i}^{(k)}R_{ij}}{\sum_{i=1}^{N_{\rm ref}}W_{1i}^{(k)2}}. \numberthis
\end{align*}
When we denote the corresponding component matrix generated by missing data with $H'$, we have
\begin{align*}
H_{1j}'^{(k+1)} &= \frac{\sum_{i=1, \mathbbm{1}_{ij} = 1}^{N_{\rm ref}}W_{1i}^{(k)}R_{ij}}{\sum_{i=1}^{N_{\rm ref}}W_{1i}^{(k)2}} \label{eq-a11-h1j}\numberthis\\
			&= H_{1j}^{(k+1)} - \frac{\sum_{i=1, \mathbbm{1}_{ij} = 0}^{N_{\rm ref}}W_{1i}^{(k)}R_{ij}}{\sum_{i=1}^{N_{\rm ref}}W_{1i}^{(k)2}}.\label{eq-a12-h1j}\numberthis
\end{align*}
Noticing $H_{1j}$ is applied to $W_1$ to approximate the $j$-th column of $R$, i.e., the $j$-th element in each flattened reference image, when the data missing rate of the $j$-th element in the reference images $R_{(\cdot)j}$ is relatively small and does not have significant variation across different pixels, i.e., 
\begin{equation}
\epsilon_{\rm pixel} \equiv \frac{\sum_{i=1, \mathbbm{1}_{ij} = 0}^{N_{\rm ref}}1}{N_{\rm ref}} \ll 1, \label{assump-a2-2}
\end{equation}
we can rewrite Equation~\eqref{eq-a12-h1j} as
\begin{equation}
H_{1j}'^{(k+1)}  = H_{1j}^{(k+1)} [1-o(1)]. \label{eq-a3-h1i}
\end{equation}

Third, we note that in the above connection between $H_{1j}'^{(k+1)}$ and $H_{1j}^{(k+1)}$ in Equation~\eqref{eq-a3-h1i}, we assumed that the coefficient matrix $W_1$ was not impacted by the missing data. We now replace $W_1$ with $W_1'$ in Equation~\eqref{eq-a11-h1j} using Equation~\eqref{eq-a9-w1i} to reflect the actual procedure, then we have
\begin{align*}
H_{1j}'^{(k+1)} &= \frac{\sum_{i=1, \mathbbm{1}_{ij} = 1}^{N_{\rm ref}}W_{1i}^{'(k)}R_{ij}}{\sum_{i=1}^{N_{\rm ref}}W_{1i}^{'(k)2}}\\
			&= \frac{\sum_{i=1, \mathbbm{1}_{ij} = 1}^{N_{\rm ref}}W_{1i}^{(k)}R_{ij}[1-o(1)]}{\sum_{i=1}^{N_{\rm ref}}W_{1i}^{(k)2}[1-o(1)]^2}\\
			&= [1+o(1)] \frac{\sum_{i=1, \mathbbm{1}_{ij} = 1}^{N_{\rm ref}}W_{1i}^{(k)}R_{ij}}{\sum_{i=1}^{N_{\rm ref}}W_{1i}^{(k)2}},
\end{align*}
where the derivation is kept up to the first order, and this fraction is the identical expression with Equation~\eqref{eq-a11-h1j}. When we follow the derivation of Equation~\eqref{eq-a3-h1i} for the above equation, and replace the small $o$ notations with $\epsilon_{\rm image}$ and $\epsilon_{\rm pixel}$ in Equations~\eqref{assump-a2-1} and \eqref{assump-a2-2}, we have
\begin{equation}
H_{1j}'^{(k+1)}  = (1+\epsilon_{\rm image}) H_{1j}^{(k+1)} (1-\epsilon_{\rm pixel}). \label{eqa-1st-order}
\end{equation}

In the above equation, we expect that the deviation between $H'$ and $H$ to be first order when either Equation~\eqref{assump-a2-1} or Equation~\eqref{assump-a2-2} dominates the fraction of missing data. Nevertheless, when the two terms are equal, i.e., $\epsilon_{\rm image} = \epsilon_{\rm pixel} = \epsilon$, Equation~\eqref{eqa-1st-order} can be rewritten as
\begin{align*}
H_{1j}'^{(k+1)}  &= (1-\epsilon^2) H_{1j}^{(k+1)}  \\
			&= [1-o^2(1)] H_{1j}^{(k+1)}, \numberthis \label{eqa-2nd-order}
\end{align*}
which is a second order deviation. As an example, this is consistent with the \citet{zhu16} observation of a few percent deviation in NMF components when $20\%$ of the data were randomly discarded in their Section 3.3, since $20\%\times20\%=4\%$. This deviation is consequently propagated to both $W$ and $H$ in Equation~\eqref{eq-a8-w1i} and Equation~\eqref{eq-a12-h1j}, respectively. As a result, we conclude that the deviation of the sNMF components induced by missing data is between the first oder and the second order. \QED

\section{sNMF Target Modeling}\label{di:model}
In this Appendix, we convert the target modeling problem for RDI, ADI, and SDI post-processing scenarios to a ``missing data'' problem in statistics. Identical to the component construction procedure in \ref{append-nmf-di}, this is achieved by assigning zero weight to the known regions containing circumstellar signals.

\subsection{Data Imputation for Missing Data}\label{di:model1}
{\bf Lemma 2} \textit{(Modeling)}:  Zero-weighted ``missing data'' region in an image can be imputed with PSF-only signals using the sNMF components. 

{\bf Proof}: In target modeling, for a given target exposure $T\in\mathbb{R}^{1\times N_{\rm pix}}$ with element-wise squared uncertainty $v\in\mathbb{R}^{1\times N_{\rm pix}}$, we use the sNMF components to model it with Equation~\eqref{eq-a3:model}. In this step, the regions containing circumstellar signals are ignored by assigning zero weights to them, and we ignore the $v$ matrix for simplicity of proof as in \ref{append-nmf-di}. With the introduction of another indicator row vector $\mathbbm{1}_T \in \mathbb{R}^{1\times N_{\rm pix}}$ as in Equation~\eqref{eq-a6:indicator-help}, the coefficient row matrix is updated as follows: 
\begin{align*}
\omega^{(k+1)} &= \omega^{(k)}\circ\frac{TH^T}{\omega^{(k)}HH^T}\\
			&\equiv \omega^{(k)}\circ\frac{(T\circ \mathbbm{1}_T)H^T}{[\omega^{(k)}H\circ \mathbbm{1}_T]H^T}\\
			&= \omega^{(k)}\circ\frac{(T\circ \mathbbm{1}_T)H^T}{[\omega^{(k)H}\circ \mathbbm{1}_T]H^T}.\numberthis
\end{align*}
The $i$-th element of $\omega$ ($i=1, 2, \cdots, n$) is thus
\begin{equation}
\omega_i^{(k+1)} = \omega_i^{(k)}\frac{\sum_{j=1}^{N_{\rm pix}}T_j \mathbbm{1}_{T, j} H_{ij}}{\sum_{j=1}^{N_{\rm pix}}\left(\omega^{(k)}H\right)_j \mathbbm{1}_{T, j} H_{ij}},
\end{equation}
where the numerator ,
\begin{align*}
\sum_{j=1}^{N_{\rm pix}}T_j \mathbbm{1}_{T, j} H_{ij} &= \sum_{j=1, \mathbbm{1}_{T,j}=1}^{N_{\rm pix}}T_j H_{ij}\mathbbm{1}_{T, j}  + \sum_{j=1, \mathbbm{1}_{T,j}=0}^{N_{\rm pix}}T_j  H_{ij} \mathbbm{1}_{T, j}\\
		&= \sum_{j=1, \mathbbm{1}_{T,j}=1}^{N_{\rm pix}}T_j H_{ij}.\numberthis\label{eq-partial-sum}
\end{align*}
Therefore, following the same argument on the relationship between Equation \eqref{eq-a8:individual} and Equation~\eqref{eq-a4:first-comp}, we conclude that the circumstellar signals do not have impact on the modeling of a target exposure, and both $\omega$ and $H$ contain the contribution from only the non-circumstellar signals.  \QED

\subsection{Coefficient Variation Induced by Missing Data}\label{di:coefs}

{\bf Theorem 2} \textit{(Imputation)}: When the signal contribution from the masked region is relatively small, for a given target and NMF basis, the coefficients for the target do not have significant variation between the non-masked and masked scenarios.

{\bf Proof}: When the NMF iteration converges, Equation~(35) in \citet{ren18} states, for component $H_i$, its coefficient $\omega_i$ has a form of
\begin{equation}
\omega_i = \frac{TH_i^T}{H_iH_i^T}\frac{1}{1+\sum_{j=1, j\neq i}^n \frac{\omega_j}{\omega_i}\frac{H_jH_i^T}{H_iH_i^T}}.\label{eq-converge-coef}
\end{equation}

Given the sparse coefficients in NMF modeling \citep{ren18}, for different components, we have two categories.

First, the coefficient is zero. Due to the non-negativity constraint in Equation~\eqref{eq-converge-coef}, the proof is trivial. 

Second, the coefficient is non-zero. If we denote the masked coefficient by $w_i'$, then
\begin{equation}
\omega_i' = \frac{(T\circ \mathbbm{1}_T)H_i^T}{(H_i\circ \mathbbm{1}_T)(H_i\circ \mathbbm{1}_T)^{T}}\frac{1}{1+\sum_{j=1, j\neq i}^n \frac{\omega_j'}{\omega_i'}\frac{(H_j\circ \mathbbm{1}_T)(H_i\circ \mathbbm{1}_T)^{T}}{(H_i\circ \mathbbm{1}_T)(H_i\circ \mathbbm{1}_T)^{T}}}.\label{eq-converge-coef2}
\end{equation}

In this category, the theorem is equivalent to proving
\begin{equation}
\left|\omega_i-\omega_i'\right| = o(\omega_i).\label{eq-corro-equiv}
\end{equation}

We first focus on the absolute difference of the first term on the righthand side of Equations~\eqref{eq-converge-coef} and \eqref{eq-converge-coef2}. When we substitute Equation~\eqref{eq-partial-sum} into the difference, we have
\begin{align*}
&\left|\frac{TH_i^T}{H_iH_i^T}-\frac{(T\circ \mathbbm{1}_T)H_i^T}{(H_i\circ \mathbbm{1}_T)(H_i\circ \mathbbm{1}_T)^{T}}\right| \numberthis \label{eq-derive-1}\\
=&\frac{TH_i^T}{H_iH_i^T}\left|1-\frac{(T\circ \mathbbm{1}_T)H_i^T}{(H_i\circ \mathbbm{1}_T)(H_i\circ \mathbbm{1}_T)^{T}}\frac{H_iH_i^T}{TH_i^T}\right| \\
=&\frac{TH_i^T}{H_iH_i^T}\left|1-\frac{H_iH_i^T}{(H_i\circ \mathbbm{1}_T)(H_i\circ \mathbbm{1}_T)^{T}}\frac{(T\circ \mathbbm{1}_T)H_i^T}{TH_i^T}\right| \\
=&\frac{TH_i^T}{H_iH_i^T}\left|1-\frac{\sum_{j=1}^{N_{\rm pix}}H_{ij}^2}{\sum_{j=1, \mathbbm{1}_{T,j}=1}^{N_{\rm pix}}H_{ij}^2}\frac{\sum_{j=1, \mathbbm{1}_{T,j}=1}^{N_{\rm pix}}T_j H_{ij}}{\sum_{j=1}^{N_{\rm pix}}T_j H_{ij}}\right| \\
=&\frac{TH_i^T}{H_iH_i^T}\left|1-\frac{1-\frac{\sum_{j=1, \mathbbm{1}_{T,j}=0}^{N_{\rm pix}}H_{ij}^2}{\sum_{j=1, \mathbbm{1}_{T,j}=1}^{N_{\rm pix}}H_{ij}^2}}{1-\frac{\sum_{j=1, \mathbbm{1}_{T,j}=0}^{N_{\rm pix}}T_j H_{ij}}{\sum_{j=1, \mathbbm{1}_{T,j}=1}^{N_{\rm pix}}T_j H_{ij}}}\right|.\numberthis\label{eq-diff-limit}
\end{align*}

When the summation fractions in Equation~\eqref{eq-diff-limit} are negligible, i.e., when
\begin{equation}\label{eq-assumption1}
\frac{{\sum_{j=1, \mathbbm{1}_{T,j}=0}^{N_{\rm pix}}H_{ij}^2}}{\sum_{j=1, \mathbbm{1}_{T,j}=1}^{N_{\rm pix}}H_{ij}^2} = o(1)
\end{equation}
and
\begin{equation}\label{eq-assumption2}
\frac{\sum_{j=1, \mathbbm{1}_{T,j}=0}^{N_{\rm pix}}T_j H_{ij}}{\sum_{j=1, \mathbbm{1}_{T,j}=1}^{N_{\rm pix}}T_j H_{ij}} = o(1),
\end{equation}
we can rewrite Equation~\eqref{eq-diff-limit} as
\begin{align*}
&\left|\frac{TH_i^T}{H_iH_i^T}-\frac{(T\circ \mathbbm{1}_T)H_i^T}{(H_i\circ \mathbbm{1}_T)(H_i\circ \mathbbm{1}_T)^{T}}\right| \\
=&\frac{TH_i^T}{H_iH_i^T} \left|1-[1-o(1)][1+o(1)] \right|\\
 =& o^2\left(\frac{TH_i^T}{H_iH_i^T}\right).\numberthis\label{eq-2o0}
\end{align*}
Similarly, following the derivation for Equation~\eqref{eq-diff-limit}, the other terms on the righthand side of Equations~\eqref{eq-converge-coef} and \eqref{eq-converge-coef2} can be proven to have negligible deviation. What is more, the proof for Equation~\eqref{eq-2o0} is obtained to the second order, therefore,
\begin{equation}
\left|\omega_i-\omega_i'\right| = o^2(\omega_i).\label{eq-2o}
\end{equation}

Combining the two categories, the proof for the theorem is complete. \QED

\paragraph{Notes} The above theorem is valid only when Equations~\eqref{eq-assumption1} and \eqref{eq-assumption2} hold. Given the fact that both equations are masked at the same region, when the target, $T$, resembles the PSF signals, Equation~\eqref{eq-assumption2} can be deduced from Equations~\eqref{eq-assumption1}. We therefore only need to demonstrate that Equation~\eqref{eq-assumption1} is valid for direct imaging data using the HR~4796A observations in Section~\ref{sec-disk}. We calculate the radial profile for the DI-sNMF components (see Figure~\ref{fig:hrcomp} for two examples), then present the squared radial profile that is divided by the sum of the squares for each component in Figure~\ref{fig:radialprofile-component-squared}. 

In Figure~\ref{fig:radialprofile-component-squared}, we observe that the highest fraction is ${\sim}10^{-3}$, and most of the fraction drops logarithmically when the radial separation increases. Based on these facts, and the fact that most of our masked pixels are far from the central region, the contribution from the masked region is expected to be smaller. For the outlier (i.e., component 3), the profile increases then reaches a plateau, which means that the entries of this component are roughly equally weighted. Equation~\eqref{eq-assumption1} therefore holds for this high contrast imaging data. Following similar arguments, for point source modeling in Section~\ref{sec-planet}, this deviation is expected to be negligible since point sources have significantly less impacts than extended structures.

In our reduction for the HR~4796A observations, we have flagged ${\sim}15\%$ of the region as ``missing data'' for data imputation. Equation~\eqref{eq-assumption1} is still valid in this case, since the overall influence measured from the variance is ${\sim}10\%$. Furthermore, the influence is expected to be significantly smaller than ${\sim}10\%$, which may even reach its squared value (i.e., ${\sim}1\%$) since Equation~\eqref{eq-2o} is a small number to the second order. Even if the influence is at ${\sim}10\%$, it will not change our results in Section~\ref{sec-disk-result}, since we expect a systematic offset due to the identical treatment of the data.
\section{Auxiliary Figures}\label{append-b}

\begin{figure}[hbt!]
\center
\includegraphics[width=.5\textwidth]{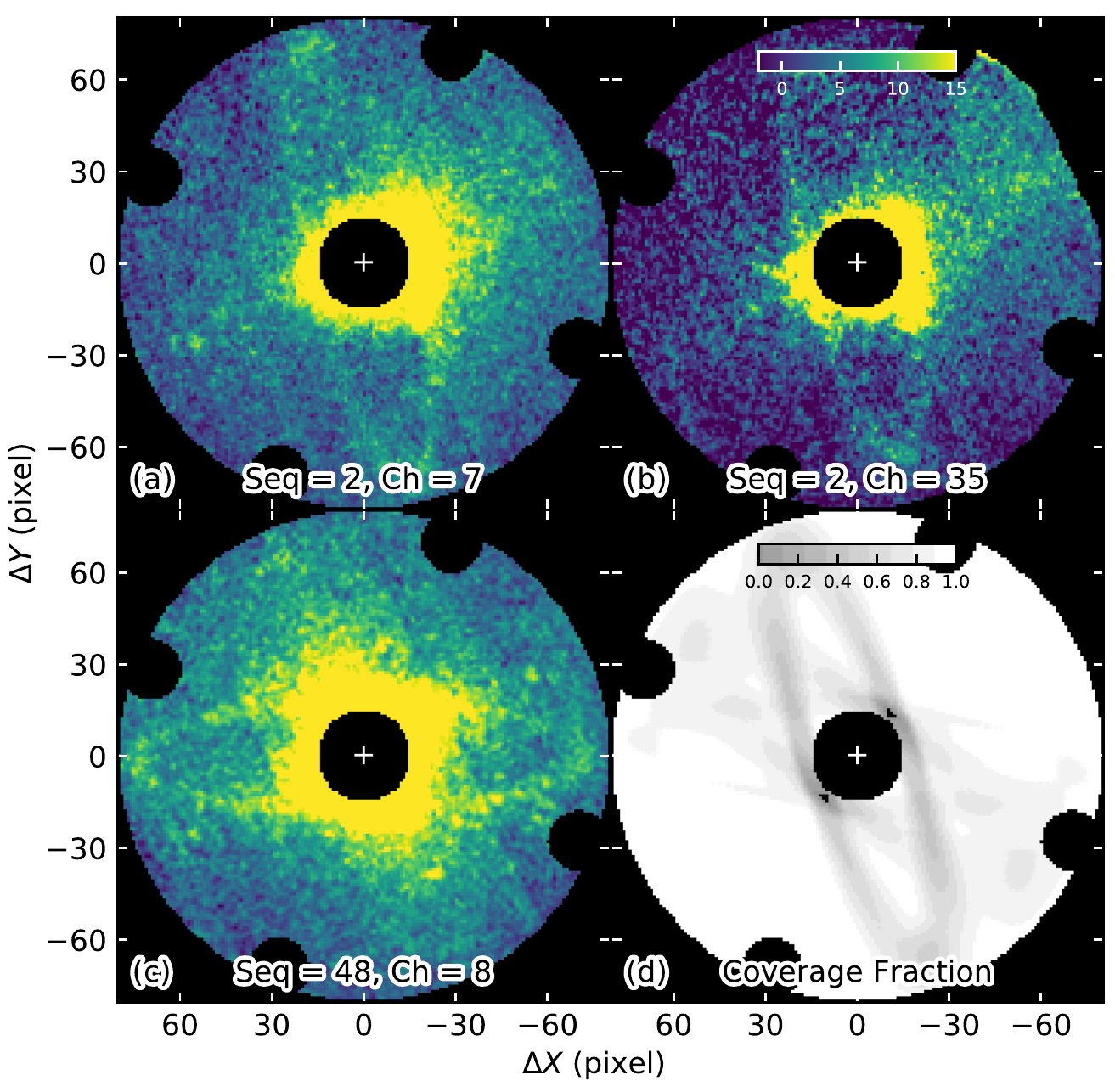}
\caption{Example spatially scaled HR~4796A $K1$-coron images in detector counts (\textbf{a}, \textbf{b}, and \textbf{c} in different sequence and/or channel numbers), and the field coverage fraction after disk exclusion (\textbf{d}). With the spatial scaling and rotating procedures (for SDI and ADI, respectively), most of the field is covered by PSF-only signals (median: $90\%$, $3\sigma$ lower limit: $5\%$). Note: (1) the exposure images share the same color bar; (2) only the pure dark regions are always excluded.}
\label{fig:hrcoverage}
\end{figure}

\begin{figure}[hbt!]
\center
\includegraphics[width=.5\textwidth]{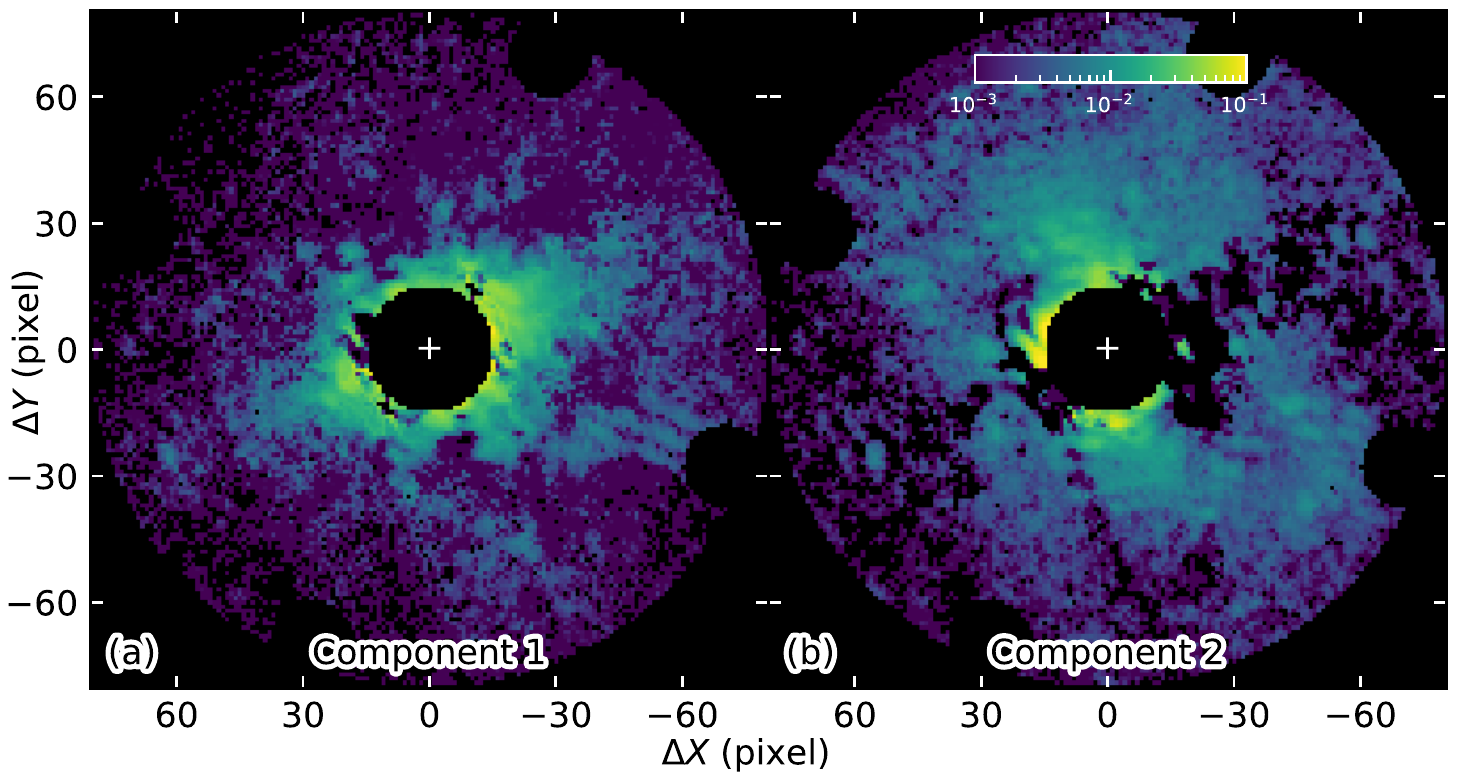}
\caption{Normalized DI-sNMF component examples. The components are constructed from the disk-excluded and channel-collapsed $240$ images of HR~4796A. No disk signal is captured in the components as expected. Note: the images share the same color bar.}
\label{fig:hrcomp}
\end{figure}

\begin{figure}[h!]
\center
\includegraphics[width=.45\textwidth]{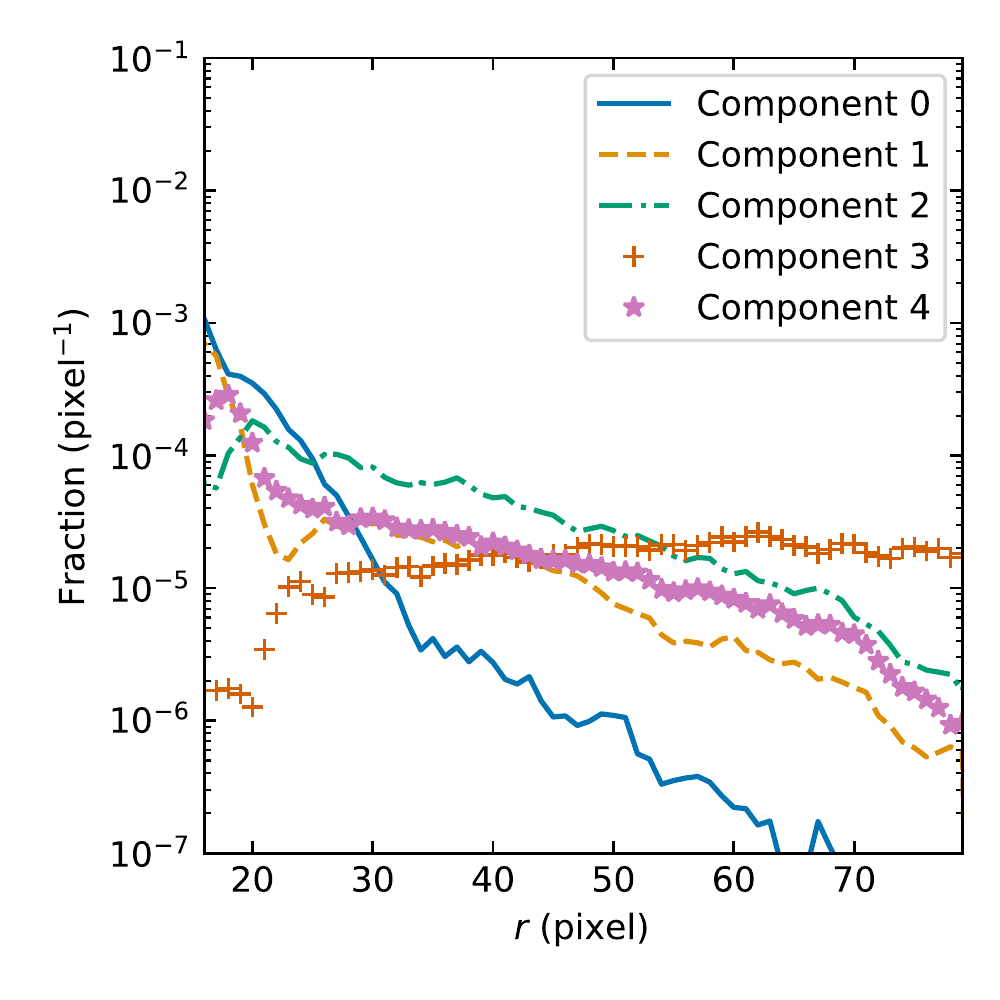}
\caption{Pixel-wise fractional squared radial profiles for typical DI-sNMF components for HR~4796A observations in this paper. For each component, the fractional squared radial profile is obtained by calculating the radial profile of the squared component, and divide that profile by the sum of the squared values of the component. For each component, the squared contribution is at most at $10^{-3}$ for one pixel, and most of the values drops logarithmically when the radial separation increases. The assumption in Equation~\eqref{eq-assumption1} is thus valid. See the notes in Appendix~\ref{di:coefs} for a detailed discussion.}
\label{fig:radialprofile-component-squared}
\end{figure}

We present the spatially scaled raw exposures for HR~4796A in Figure~\ref{fig:hrcoverage}, the sNMF components constructed from the ``missing data'' approach in Figure~\ref{fig:hrcomp}, and the fractional variance radial profiles for typical components of the HR~4796A observation in Figure~\ref{fig:radialprofile-component-squared}.

\bibliography{refs}
\end{CJK*}
\end{document}